\documentstyle[preprint,amstex,aps,eqsecnum,epsfig]{revtex}
\tighten
\def\E{{\cal E}}                            % F
\def\N{{\cal N}}                            % N
\def\T{{\cal T}}                            % T 
\def\V{{\cal V}}                            % V 
\def\L{\Lambda}

\def\l{\lambda}

\def\s{\sigma}
\def\om{\omega}

\def\t{\theta}

\def\ZZ{{\Bbb Z }}
\def\bds#1{\boldsymbol{#1}}
\def\lowmp{\lower.11em\hbox{${\scriptstyle\mp}$}}
\def\intf{\int_{-\infty}^{+\infty}}

\def\abs#1{\left| #1\right|}                % abs
\def\VEV#1{\left\langle #1 \right\rangle}   % Vacuum Expect. Value
\def\bra#1{\left\langle #1\right|}               % bra
\def\ket#1{\left| #1\right\rangle}               % ket
\def\der#1#2{{d#1\over d#2}}
\def\pdif#1#2{{\partial #1 \over \partial #2}}
\def\braket#1#2{\left\langle #1 \right| \left. #2 \right\rangle}
\def\VEV#1{\left\langle #1 \right\rangle}
\def\vev#1{\langle #1 \rangle}
\def\mbare{m^2_{\text{b}}}
\def\lbare{\lambda_{\text{b}}}

\begin{document}
\preprint{ IFUM 651/FT-99 \,\, Bicocca-FT-99-41\bigskip}
%\draft
\title{\bf An improved time--dependent Hartree--Fock approach for scalar 
$\phi^4$ QFT}
\author{{\bf C. Destri $^{(a,b)}$}  and {\bf E. Manfredini $^{(b)}$\bigskip}}

\bigskip

\address{
(a) Dipartimento di Fisica G. Occhialini, \\
   Universit\`a di Milano--Bicocca  and INFN, sezione di Milano$^{ 1,2}$
    \\
(b)  Dipartimento di Fisica,  Universit\`a di Milano \\ 
     and INFN, sezione di Milano$^{ 1,2}$
}
\footnotetext{
$^1$mail address: Dipartimento di Fisica, Via Celoria 16, 20133 Milano,
ITALIA.}
\footnotetext{
$^2$e-mail: claudio.destri@@mi.infn.it, emanuele.manfredini@@mi.infn.it
}
\date{November 1999}
\maketitle
\begin{abstract}
The $\lambda \phi^4$ model in a finite volume is studied within a
non--gaussian Hartree--Fock approximation (tdHF) both at equilibrium
and out of equilibrium, with particular attention to the structure of
the ground state and of certain dynamical features in the broken
symmetry phase. The mean--field coupled time--dependent Schroedinger
equations for the modes of the scalar field are derived and the
suitable procedure to renormalize them is outlined. A further
controlled gaussian approximation of our tdHF approach is used in
order to study the dynamical evolution of the system from
non--equilibrium initial conditions characterized by a uniform
condensate. We find that, during the slow rolling down, the
long--wavelength quantum fluctuations do not grow to a macroscopic
size but do scale with the linear size of the system, in accordance
with similar results valid for the large $N$ approximation of the
$O(N)$ model. This behavior undermines in a precise way the gaussian
approximation within our tdHF approach, which therefore appears as a
viable mean to correct an unlikely feature of the standard HF
factorization scheme, such as the so--called ``stopping at the spinodal
line'' of the quantum fluctuations. We also study the dynamics of the
system in infinite volume with particular attention to the asymptotic
evolution in the broken symmetry phase. We are able to show that the
fixed points of the evolution cover at most the classically metastable
part of the static effective potential.
\end{abstract}

\newpage

\section{Introduction}\label{int}
A great effort has been devoted in the last few years in order to
develop a deeper qualitative and quantitative understanding of
systems described by interacting quantum fields out of
equilibrium. There is a class of physical problems that requires the
consistent treatment of time dependent mean--fields in interaction
with their own quantum or thermal fluctuations. We may mention, among
others, the problem of reheating of the universe after the
inflationary era of exponential growth and cooling, and the time
evolution of the scalar order parameter through the chiral phase
transition, soon to be probed in the forthcoming heavy--ion experiments
at CERN--SPS, BNL--RHIC and CERN--LHC. In these situations, a detailed
description of the time--dependent dynamics is necessary to calculate
the non--equilibrium properties of the system. Indeed, the
development of practical general techniques and the advent of faster
and cheaper computers have made possible the discovery of novel and
unexpected phenomena, ranging from dissipative processes via particle
production to novel aspects of symmetry breaking
\cite{devega2,chkm,relax,tsunami}.

From the technical point of view, it should be pointed out, first of
all, that a perturbative treatment of this dynamical problem is
meaningful only when the early time evolution is considered. The
presence of parametric resonant bands or spinodal instabilities (in
the case, respectively, of unbroken or spontaneously broken symmetries)
rapidly turns the dynamics completely non--linear and
non--perturbative. Thus, the asymptotic evolution at late time can be
consistently studied only if approximate non--perturbative methods are
applied to the problem \cite{devega2}.

Quite recently one of these schemes, namely the large $N$ expansion at
leading order (LN) \cite{largen_exp,chkmp}, has been used in order to
clarify some dynamical aspects of the $\phi^4$ theory in $3$ spatial
dimensions, reaching the conclusion that the non--perturbative and
non--linear evolution of the system might eventually produce the onset
of a form of non--equilibrium Bose--Einstein condensation (BEC) of the
long--wavelength Goldstone bosons usually present in the broken
symmetry phase \cite{relax,tsunami,bdvhs}. Another very interesting
result in \cite{bdvhs} concerns the dynamical Maxwell construction,
which reproduces the flat region of the effective potential in case of
broken symmetry as asymptotic fixed points of the background
evolution.

In a companion work \cite{finvN} we have addressed the question of
whether a standard BEC could actually take place as time goes on, by
putting the system in a finite volume (a periodic box of size $L$) and
carefully studying the volume dependence of out--of--equilibrium
features in the broken symmetry phase. We summarize here the main
result contained in \cite{finvN}. The numerical solution shows the
presence of a time scale $\tau_L$, proportional to the linear size $L$
of the system, at which finite volume effects start to manifest, with
the remarkable consequence that the zero-mode quantum fluctuations
cannot grow macroscopically large if they start with microscopic
initial conditions. In fact, the size of low--lying widths at time
$\tau_L$ is of order $L$, to be compared to order $L^{3/2}$ for the
case of standard BEC. In other words we confirmed that the linear
growth of the zero mode width, as found also by the authors of
\cite{relax,tsunami,bdvhs}, really signals the onset of a novel form
of dynamical BEC, quite different from the standard one described by
equilibrium finite--temperature field theory. This interpretation is
reinforced by the characteristics of the long--wavelength
fluctuations' spectrum.

Since after all the large $N$ approximation is equivalent to a
Gaussian ansatz for the time--dependent density matrix of the system
\cite{chkm,yaffe}, one might still envisage a scenario in which, while
gaussian fluctuations would stay microscopic, non--gaussian
fluctuations would grow in time to a macroscopic size, leading to an
occupation number for the zero mode proportional to the volume $L^3$
of the system. Therefore, in order to go beyond the gaussian
approximation, we will consider in this work a time--dependent HF
approach capable in principle of describing the dynamics of some
non--gaussian fluctuations of a single scalar field with $\phi^4$
interaction.

Before going into the details of the analysis, let us briefly
summarize the main limitations and the most remarkable results of the
study of a scalar field out of equilibrium within the gaussian HF
scheme \cite{devega2,devega3,erice,devega7}. First of all, this scheme
has the advantage of going beyond perturbation theory, in the sense
that the (numerical) solution of the evolution equations will contain
arbitrary powers of the coupling constant, corresponding to a
non--trivial resummation of the perturbative series. For this reason,
the method is able to take into account the quantum back--reaction on
the fluctuations themselves, which shuts off their early exponential
growth. This is achieved by the standard HF factorization of the
quartic interaction, yielding a {\em time dependent}
self--consistently determined mass term, which stabilizes the modes
perturbatively unstable. The detailed numerical solution of the
resulting dynamical equations clearly shows the dissipation associated
with particle production, as a result of either parametric
amplification in case of unbroken symmetry or spinodal instabilities
in case of broken symmetry, as well as the shut off mechanism outlined
above.

However, the standard HF method is really not controllable in the case
of a single scalar field, while it becomes exact only in the $N \to
\infty$ limit. Moreover, previous approaches to the dynamics in this
approximation scheme had the unlikely feature of maintaining a weak
(logarithmic) cut--off dependence on the renormalized equations of
motion of the order parameter and the mode functions \cite{devega2}.

In this article we consider the case of a single scalar field
(i.e. $N=1$). With the aim of studying the dynamics of the model with
the inclusion of some non--gaussian contributions, we introduce an
improved time--dependent Hartree--Fock approach. Even if it is still
based on a factorized trial wavefunction(al), it has the merit to keep
the quartic interaction diagonal in momentum space, explicitly in the
hamiltonians governing the evolution of each mode of the field. In
this framework, issues like the static spontaneous symmetry breaking
can be better understood, and the further gaussian approximation
needed to study the dynamics can be better controlled. In particular,
questions like out--of--equilibrium ``quantum phase ordering'' and ``dynamical
Bose--Einstein condensation'' can be properly posed and answered
within a verifiable approximation.

We also perform a detailed study of the asymptotic dynamics in
infinite volume, with the aim of clarifying the issue of Maxwell
construction in this approximation scheme. In fact, in the $O(N) \,
\Phi ^4$ model at leading order, the asymptotic dynamical evolution of
the mean field completely covers the spinodal region of the classical
potential, which coincides with the flatness region of the effective
potential. This is what is called {\em dynamical Maxwell construction}
\cite{bdvhs}. When we use the HF approximation for the case of $N=1$,
we find that the spinodal region and the flatness region are different
and the question arise of whether a full or partial dynamical Maxwell
construction still takes place.

In section \ref{cft} we set up the model in finite volume, defining
all the relevant notations and the quantum representation we will be
using to study the evolution of the system.

We introduce in section \ref{tdhf} our improved time--dependent
Hartree--Fock (tdHF) approximation, which generalizes the standard
gaussian self-consistent approach \cite{tdHF} to non--gaussian
wave--functionals; we then derive the mean--field coupled
time--dependent Schroedinger equations for the modes of the scalar
field, under the assumption of a uniform condensate, see eqs
(\ref{Schroedinger}), (\ref{H_k}) and (\ref{omvev}). A significant
difference with respect to previous tdHF approaches \cite{devega2}
concerns the renormalization of ultraviolet divergences. In fact, by
means of a single substitution of the bare coupling constant $\lbare$
with the renormalized one $\l$ in the Hartree--Fock hamiltonian, we
obtain cut-off independent equations (apart from corrections in
inverse powers, which are there due to the Landau pole). The
substitution is introduced by hand, but is justified by simple
diagrammatic considerations.

One advantage of not restricting a priori the self-consistent HF
approximation to gaussian wave--functionals, is in the possibility of
a better description of the vacuum structure in case of broken
symmetry. In fact we can show quite explicitly that, in any finite
volume, in the ground state the zero--mode of $\phi$ field is
concentrated around the two vacua of the broken symmetry, driving the
probability distribution for any sufficiently wide smearing of the
field into a two peaks shape. This is indeed what one would
intuitively expect in case of symmetry breaking. On the other hand
none of this appears in a dynamical evolution that starts from a
distribution localized around a single value of the field in the
spinodal region, confirming what already seen in the large $N$
approach \cite{finvN}. More precisely, within a further controlled
gaussian approximation of our tdHF approach, one observes that
initially microscopic quantum fluctuations never becomes macroscopic,
suggesting that also non--gaussian fluctuations cannot reach
macroscopic sizes.  As a simple confirmation of this fact, consider
the completely symmetric initial conditions
$\vev{\phi}=\vev{\dot\phi}=0$ for the background: in this case we find
that the dynamical equations for initially gaussian field fluctuations
are identical to those of large $N$ (apart for a rescaling of the
coupling constant by a factor of three; cfr. ref. \cite{finvN}), so
that we observe the same asymptotic vanishing of the effective
mass. However, this time no interpretation in terms of Goldstone
theorem is possible, since the broken symmetry is discrete; rather, if
the width of the zero--mode were allowed to evolve into a macroscopic
size, then the effective mass would tend to a positive value, since
the mass in case of discrete symmetry breaking is indeed larger than
zero.

Anyway, also in the gaussian HF approach, we do find a whole class of
cases which exhibit the time scale $\tau_L$. At that time, finite
volume effects start to manifest and the size of the low--lying widths
is of order $L$. We then discuss why this undermines the
self--consistency of the gaussian approximation, imposing the need of
further study, both analytical and numerical.

In section \ref{ep} we study the asymptotic evolution in the broken
symmetry phase, in infinite volume, when the expectation value starts
within the region between the two minima of the potential. We are able
to show by precise numerical simulations, that the fixed points of the
background evolution do not cover the static flat region
completely. On the contrary, the spinodal region seems to
be absolutely forbidden for the late time values of the mean
field. Thus, as far as the asymptotic evolution is concerned, our
numerical results lead to the following conclusions. We can
distinguish the points lying between the two minima in a fashion
reminiscent of the static classification: first, the values satisfying
the property $v/\sqrt{3} < \abs{\bar\phi _{\infty}} \leq v$ are
metastable points, in the sense that they are fixed points of the
background evolution, no matter which initial condition comprised in
the interval $\left( -v , v \right)$ we choose for the expectation
value $\bar\phi$; secondly, the points included in the interval $0 <
\abs{\bar\phi _{\infty}} < v/\sqrt{3}$ are unstable points, because if
the mean field starts from one of them, after an early slow rolling
down, it starts to oscillate with decreasing amplitude around a point
inside the classical metastable interval. Obviously, $\bar\phi=v$ is
the point of stable equilibrium, and $\bar\phi=0$ is a point of
unstable equilibrium. Actually, it should be noted that our data do
not allow a precise determination of the border between the dynamical
unstable and metastable regions; thus, the number we give here should
be looked at as an educated guess inspired by the analogous static
classification and based on considerations about the solutions of the
gap equation [see eq. (\ref{newgap})]

Finally, in section \ref{conclusion} we give a brief summary of the results
presented in this article and we outline some interesting open
questions that need more work before being answered properly.

\section{Cutoff field theory}\label{cft}

Let us consider the scalar field operator $\phi$ and its canonically
conjugated momentum $\pi$ in a $D-$dimensional periodic box of size
$L$ and write their Fourier expansion as customary
\begin{equation*}
	\phi(x)=L^{-D/2}\sum_k \phi_k\,e^{ik\cdot x} 
	\;,\quad \phi_k^\dag = \phi_{-k}
\end{equation*}
\begin{equation*}
	\pi(x)=L^{-D/2}\sum_k \pi_k\,e^{ik\cdot x} 
	\;,\quad \pi_k^\dag = \pi_{-k}
\end{equation*}
with the wavevectors $k$ naturally quantized: $k=(2\pi/L)n$, $n\in\ZZ^D$.

The canonical commutation rules are
$[\phi_k\,,\pi_{-k'}]=i\delta^{(D)}_{kk'}$, as usual. The introduction
of a finite volume should be regarded as a regularization of the
infrared properties of the model, which allows to ``count'' the
different field modes and is needed especially in the case of broken
symmetry.

To keep control also on the ultraviolet behavior and manage to handle
the renormalization procedure properly, we restrict the sums over
wavevectors to the points lying within the $D-$dimensional sphere of
radius $\Lambda$, that is $k^2\le\Lambda^2$, with $\N=\Lambda L/2\pi$
some large integer. Till both the cut--offs remain finite, we have
reduced the original field--theoretical problem to a
quantum--mechanical framework with finitely many (of order $\N^{D-1}$)
degrees of freedom. 

The $\phi^4$ Hamiltonian is
\begin{equation*}
\begin{split}
  H &=\dfrac12\int d^Dx\,\left[\pi^2 + (\partial\phi)^2 + \mbare\,\phi^2
	+ \lbare\, \phi^4 \right] = \\
    &=\dfrac12\sum_k\left[\pi_k\pi_{-k} +(k^2+ \mbare)\,\phi_k\phi_{-k}\right]
	+\dfrac\l4 \sum_{k_1,k_2,k_3,k_4}\phi_{k_1}\phi_{k_2}
	 \phi_{k_3}\phi_{k_4}\,\delta^{(D)}_{k_1+k_2+k_3+k_4,0}
\end{split}
\end{equation*}
where $\mbare$ and $\lbare$ are the bare parameters and depend on the
UV cutoff $\Lambda$ in such a way to guarantee a finite limit
$\Lambda\to\infty$ for all observable quantities. It should be noted
here that, being the theory trivial \cite{wilson} (as is manifest in
the resummed one--loop approximation due to the Landau pole) the
ultraviolet cut--off should be kept finite and much smaller than the
renormalon singularity. In this case, we must regard the $\phi^4$
model as an effective low--energy theory (here low--energy means
practically all energies below Planck's scale, due to the large value
of the Landau pole for renormalized coupling constants of order one or
less).

We shall work in the wavefunction representation where
$\braket{\varphi}\Psi=\Psi(\varphi)$ and
\begin{equation*}
	(\phi_0\Psi)(\varphi)=\varphi_0 \Psi(\varphi)
	  \;,\quad 
	(\pi_0\Psi)(\varphi)= -i\pdif{}{\varphi_0}\Psi(\varphi)
\end{equation*}
while for $k>0$ (in lexicographic sense)
\begin{equation*}
	(\phi_{\pm k}\Psi)(\varphi)=\dfrac1{\sqrt2}\left(
	\varphi_k\pm i\,\varphi_{-k}\right)\Psi(\varphi)
	  \;,\quad 
	(\pi_{\pm k}\Psi)(\varphi)= \dfrac1{\sqrt2} \left(-i
	\pdif{}{\varphi_k}\pm \pdif{}{\varphi_{-k}}\right)\Psi(\varphi)
\end{equation*}
Notice that by construction the variables $\varphi_k$ are all real.

In practice, the problem of studying the dynamics of the $\phi^4$
field out of equilibrium consists now in trying to solve the
time-dependent Schroedinger equation given an initial wavefunction
$\Psi(\varphi,t=0)$ that describes a state of the field far away from
the vacuum. This approach could be very well generalized in a
straightforward way to mixtures described by density matrices, as
done, for instance, in \cite{devega3,hkmp,bdv}. Here we shall restrict
to pure states, for sake of simplicity and because all relevant
aspects of the problem are already present in this case.

We shall consider here the time-dependent Hartree--Fock (tdHF)
approach (an improved version with respect to what is presented, for
instance, in \cite{tdHF}), being the large $N$ expansion to leading
order treated in another work \cite{finvN}. In fact these two methods
are very closely related (see, for instance in
\cite{inomogeneo}). However, before passing to any approximation, we
would like to stress that the following rigorous result can be
immediately established in this model with both UV and IR cutoffs.

\subsection{A rigorous result: the effective potential is convex}
\label{nogo}

This is a well known fact in statistical mechanics, being directly
related to stability requirements. It would therefore hold also for
the field theory in the Euclidean functional formulation. In our
quantum--mechanical context we may proceed as follow. Suppose the
field $\phi$ is coupled to a uniform external source $J$.  Then the
ground state energy $E_0(J)$ is a concave function of $J$, as can be
inferred from the negativity of the second order term in $\Delta J$ of
perturbation around any chosen value of $J$. Moreover, $E_0(J)$ is
analytic in a finite neighborhood of $J=0$, since $J\phi$ is a
perturbation ``small'' compared to the quadratic and quartic terms of
the Hamiltonian. As a consequence, this effective potential
$V_{\text{eff}}(\bar\phi)=E_0(J)-J\bar\phi$,
$\bar\phi=E_0'(J)=\vev{\phi}_0$, that is the Legendre transform of
$E_0(J)$, is a convex analytic function in a finite neighborhood of
$\bar\phi=0$.  In the infrared limit $L\to\infty$, $E_0(J)$ might
develop a singularity in $J=0$ and $V_{\text{eff}}(\bar\phi)$ might
flatten around $\bar\phi=0$. Of course this possibility would apply in
case of spontaneous symmetry breaking, that is for a double--well
classical potential. This is a subtle and important point that will
play a crucial role later on, even if the effective potential is
relevant for the static properties of the model rather than the
dynamical evolution out of equilibrium that interests us here. In fact
such evolution is governed by the CTP effective action
\cite{schwinger,zshy} and one might expect that, although non--local
in time, it asymptotically reduces to a multiple of the effective
potential for trajectories of $\bar\phi(t)$ with a fixed point at
infinite time. In such case there should exist a one--to--one
correspondence between fixed points and minima of the effective
potential.

\section{Time-dependent Hartree--Fock}\label{tdhf}

In order to follow the time evolution of the non--gaussian quantum
fluctuations we consider in this section a time--dependent HF
approximation capable in principle of describing the dynamics of
non--gaussian fluctuations of a single scalar field with $\phi^4$
interaction.

We examine in this work only states in which the scalar field has a
uniform, albeit possibly time--dependent expectation value.  In a tdHF
approach we may then start from a wavefuction of the factorized form
(which would be exact for free fields)
\begin{equation}\label{wf}
	\Psi(\varphi)=\psi_0(\varphi_0)
		\prod_{k>0} \psi_k(\varphi_k,\varphi_{-k})
\end{equation}
The dependence of $\psi_k$ on its two arguments cannot be assumed to
factorize in general since space translations act as $SO(2)$ rotations
on $\varphi_k$ and $\varphi_{-k}$ (hence in case of translation
invariance $\psi_k$ depends only on $\varphi_k^2+\varphi_{-k}^2$).
The approximation consists in assuming this form as valid at all times
and imposing the stationarity condition on the action
\begin{equation}\label{varpri}
	\delta \int dt\, \vev{i\partial_t-H}=0 \;,\quad
	\vev{\cdot} \equiv \bra{\Psi(t)}\cdot\ket{\Psi(t)}
\end{equation}
with respect to variations of the functions $\psi_k$. To enforce a
uniform expectation value of $\phi$ we should add a Lagrange
multiplier term linear in the single modes expectations
$\vev{\varphi_k}$ for $k\neq 0$. The multiplier is then fixed at the
end to obtain $\vev{\varphi_k}=0$ for all $k\neq 0$. Actually one may
verify that this is equivalent to the simpler approach in which
$\vev{\varphi_k}$ is set to vanish for all $k\neq 0$ before any
variation. Then the only non trivial expectation value in the
Hamiltonian, namely that of the quartic term, assumes the form
\begin{equation}\label{vevphi4}
\begin{split}
	\int d^Dx\, \vev{\phi(x)^4} = & \frac1{L^D}
	\left[ \vev{\varphi_0^4}-3\vev{\varphi_0^2}^2 \right] +
	\frac3{2L^D} \sum_{k>0} \left[
	\vev{(\varphi_k^2+\varphi_{-k}^2)^2}-2\left(\vev{\varphi_k^2}
	+\vev{\varphi_{-k}^2}\right)^2 \right] \\ &+ 
	\frac3{L^D}\left(\sum_k\vev{\varphi_k^2}\right)^2 
\end{split}
\end{equation}
Notice that the terms in the first row would cancel completely out for
gaussian wavefunctions $\psi_k$ with zero mean value. The last term,
where the sum extends to all wavevectors $k$, corresponds instead to
the standard mean field replacement $\vev{\phi^4}\to
3\vev{\phi^2}^2$. The total energy of our trial state now reads
\begin{equation}\label{energy}
	E=\vev{H} =\dfrac12\sum_k \VEV{\pdif{^2}{\varphi_k^2}+
	(k^2+\mbare)\varphi_k^2} + \dfrac\lbare{4}\int d^Dx\, \vev{\phi(x)^4}
\end{equation}
and from the variational principle (\ref{varpri}) we obtain a set
of simple Schroedinger equations
\begin{equation}\label{Schroedinger}
	i\partial_t\psi_k = H_k \psi_k 
\end{equation}
\vskip -.4truecm
\begin{equation}\label{H_k}
\begin{split}
  H_0 &=-\dfrac12\pdif{^2}{\varphi_0^2}+\dfrac12 \om_0^2 \varphi_0^2
	+\dfrac{\lbare}{4L^D}\varphi_0^4   \\
  H_k &=-\dfrac12\left(\pdif{^2}{\varphi_k^2}+\pdif{^2}{\varphi_{-k}^2}\right)
	+\dfrac12 \om_k^2 (\varphi_k^2 + \varphi_{-k}^2)
	+\dfrac{3\lbare}{8L^D}\left(\varphi_k^2+\varphi_{-k}^2\right)^2 \\
\end{split}
\end{equation}
which are coupled in a mean--field way only through 
\begin{equation}\label{omvev}
	\om_k^2 = k^2+\mbare +3\lbare \Sigma_k  \;,\quad
	\Sigma_k= \dfrac1{L^D}\sum_{{q^2\le\Lambda^2}\atop
	{q\neq k,-k}}\vev{\varphi_q^2}
\end{equation}
and define the HF time evolution for the theory. By construction this
evolution conserves the total energy $E$ of eq. (\ref{energy}).

It should be stressed that in this particular tdHF approximation,
beside the mean--field back--reaction term $\Sigma_k$ of all other
modes on $\om_k^2$, we keep also the contribution of the {\em
diagonal} scattering through the diagonal quartic terms. In fact this
is why $\Sigma_k$ has no contribution from the $k-$mode itself: in a
gaussian approximation for the trial wavefunctions $\psi_k$ the
Hamiltonians $H_k$ would turn out to be harmonic, the quartic terms
being absent in favor of a complete back--reaction
\begin{equation}\label{Sigma}
	\Sigma = \Sigma_k + \dfrac{\vev{\varphi_k^2}+
	\vev{\varphi_{-k}^2}}{L^D} = \frac1{L^D}\sum_k\vev{\varphi_k^2}
\end{equation}
Of course the quartic self--interaction of the modes as well as the
difference between $\Sigma$ and $\Sigma_k$ are suppressed by a volume
effect and could be neglected in the infrared limit, provided all
wavefunctions $\psi_k$ stays concentrated on mode amplitudes
$\varphi_k$ of order smaller than $L^{D/2}$.  This is the
typical situation when all modes remain microscopic and the volume in
the denominators is compensated only through the summation over a
number of modes proportional to the volume itself, so that in the
limit $L\to\infty$ sums are replaced by integrals
\begin{equation*}
	\Sigma_k \to \Sigma \to \int_{k^2\le\Lambda^2} 
	\dfrac{d^Dk}{(2\pi)^D} \vev{\varphi_k^2}
\end{equation*}
Indeed we shall apply this picture to all modes with $k\neq 0$, while
we do expect exceptions for the zero--mode wavefunction $\psi_0$.

The treatment of ultraviolet divergences requires particular care,
since the HF approximation typically messes things up (see, for
instance, \cite{onhartree}). Following the same login
of the large $N$ approximation \cite{devega2,finvN,chkmp}, we could
take as renormalization condition the requirement that the frequencies
$\om_k^2$ are independent of $\Lambda$, assuming that $\mbare$ and
$\lbare$ are functions of $\Lambda$ itself and of renormalized
$\Lambda-$independent parameters $m^2$ and $\l$ such that
\begin{equation}\label{renomvev}
	\om_k^2 = k^2 +m^2 +3\l\left[\Sigma_k\right]_{\text{finite}}
\end{equation}
where by $[.]_{\text{finite}}$ we mean the (scheme--dependent) finite
part of some possibly ultraviolet divergent quantity.  Unfortunately
this would not be enough to make the spectrum of energy differences
cutoff--independent, because of the bare coupling constant $\lbare$ in
front of the quartic terms in $H_k$ and the difference between
$\Sigma$ and $\Sigma_k$ [such problem does not exist in large $N$
because that is a purely gaussian approximation].  Again this would
not be a problem whenever these terms become negligible as
$L\to\infty$. At any rate, to be ready to handle the cases when this
is not actually true and to define an ultraviolet--finite model also
at finite volume, we shall by hand modify eq. (\ref{vevphi4}) as
follows:
\begin{equation}\label{subst}
\begin{split}
	\lbare\int d^Dx\, \vev{\phi(x)^4} = & \l L^{-D}\left\{
	\vev{\varphi_0^4}-3\vev{\varphi_0^2}^2 + \tfrac32
	\sum\limits_{k>0} \left[
	\vev{(\varphi_k^2+\varphi_{-k}^2)^2}-2\left(\vev{\varphi_k^2}
	+\vev{\varphi_{-k}^2}\right)^2 \right]\right\} \\ &+ 
	3\lbare \,L^D\Sigma^2 
\end{split}
\end{equation}
We keep the bare coupling constant in front of the term containing
$\Sigma^2$ because that part of the hamiltonian is properly
renormalized by means of the usual {\em cactus} resummation
\cite{cactus} which corresponds to the standard HF approximation. On
the other hand, within the same approximation, it is not possible to
renormalize the part in curly brackets of the equation above, because
of the factorized form (\ref{wf}) that we have assumed for the
wavefunction of the system. In fact, the $4-$legs vertices in the
curly brackets are diagonal in momentum space; at higher order in the
loop expansion, when we contract two or more vertices of this type, no
sum over internal loop momenta is produced, so that all higher order
perturbation terms are suppressed by volume effects. However, we know
that in the complete theory, the wavefunction is not factorized and
loops contain all values of momentum. This suggests that, in order to
get a finite hamiltonian, we need to introduce in the definition of
our model some extra resummation of Feynmann diagrams, that is not
automatically contained in this self--consistent HF approach. The only
choice consistent with the cactus resummation performed in the
two--point function by the HF scheme is the resummation of the
$1$-loop {\em fish} diagram in the four--point function. This amounts
to the change from $\lbare $ to $\l$ and it is enough to guarantee the
ultraviolet finiteness of the hamiltonian through the redefinition
\begin{equation}\label{H_k'}
	H_0 \to H_0+\dfrac{\l-\lbare}{4L^D}\varphi_0^4 \;,\quad
	H_k \to H_k +\dfrac{
	3(\l-\lbare)}{8L^D}\left(\varphi_k^2+\varphi_{-k}^2\right)^2
\end{equation}
At the same time the frequencies are now related to the widths
$\vev{\varphi_{-k}^2}$ by
\begin{equation}\label{omvev2}
\begin{split}
	\om_k^2 &= k^2+ M^2 - 3\l\, L^{-D}(\vev{\varphi_k^2}+
	\vev{\varphi_{-k}^2}) \;,\quad k>0 \\ M^2 &\equiv 
	\om_0^2 + 3\l\,L^{-D}\vev{\varphi_0^2}= \mbare +3\lbare \Sigma 
\end{split}
\end{equation}
Apart for $O(L^{-D})$ corrections, $M$ plays the role of
time--dependent mass for modes with $k\neq 0$, in the harmonic
approximation.
  
In this new setup the conserved energy reads
\begin{equation}\label{menergy}
	E =\sum_{k\ge 0}\vev{H_k} -\tfrac34 \lbare\,L^D\,\Sigma^2 +
	\tfrac34\l\,L^{-D}\left[\vev{\varphi_0^2}^2 + \sum\limits_{k>0}
	\left(\vev{\varphi_k^2}+\vev{\varphi_{-k}^2}\right)^2\right]
\end{equation}
Since the gap--like equations (\ref{omvev2}) are state--dependent, we
have to perform the renormalization first for some reference quantum
state, that is for some specific collection of wavefunctions $\psi_k$;
as soon as $\mbare$ and $\lbare$ are determined as functions
$\Lambda$, ultraviolet finiteness will hold for the entire class of
states with the same ultraviolet properties of the reference
state. Then an obvious consistency check for our HF approximation is
that this class is closed under time evolution.

Rather than a single state, we choose as reference the family of
gaussian states parametrized by the uniform expectation value
$\vev{\phi(x)}=L^{-D/2}\vev{\varphi_0}=\bar\phi$ (recall that we have
$\vev{\varphi_k}=0$ when $k\neq0$ by assumption) and such that the HF
energy $E$ is as small as possible for fixed $\bar\phi$.  Then, apart
from a translation by $L^{D/2}\bar\phi$ on $\varphi_0$, these gaussian
$\psi_k$ are ground state eigenfunctions of the harmonic Hamiltonians
obtained from $H_k$ by dropping the quartic terms. Because of the
$k^2$ in the frequencies we expect these gaussian states to dominate
in the ultraviolet limit also at finite volume (as discussed above
they should dominate in the infinite--volume limit for any
$k\neq0$). Moreover, since now
\begin{equation}\label{gauss}
	\vev{\varphi_0^2} = L^D\bar\phi^2+ \dfrac1{2\om_0} \;,\quad 
	\vev{\varphi_{\pm k}^2} = \dfrac1{2\om_k} \;,\quad k\neq 0
\end{equation}
the relation (\ref{omvev2}) between frequencies and widths turn into the
single gap equation
\begin{equation}\label{gap}
	M^2 =\mbare+3\lbare\left(\bar\phi^2
	+\dfrac1{2L^D}\sum_{q^2\le\Lambda^2}\dfrac1{\sqrt{k^2+M^2}}\right)
\end{equation}
fixing the self-consistent value of $M$ as a function of $\bar\phi$.
It should be stressed that (\ref{omvev2}) turns through
eq. (\ref{gauss}) into the gap equation only because of the
requirement of energy minimization. Generic $\psi_k$, regarded as
initial conditions for the Schroedinger equations (\ref{Schroedinger}),
are in principle not subject to any gap equation.

The treatment now follows closely that in the large $N$ approximation
\cite{finvN}, the only difference being in the value of the coupling,
now three times larger.  In fact, in case of $O(N)$ symmetry, the
quantum fluctuations over a given background
$\vev{\bds\phi(x)}=\bar{\bds\phi}$ decompose for each $k$ into one
longitudinal mode, parallel to $\bar{\bds\phi}$, and $N-1$ transverse
modes orthogonal to it; by boson combinatorics the longitudinal mode
couples to $\bar{\bds\phi}$ with strength $3\lbare/N$ and decouple in
the $N\to\infty$ limit, while the transverse modes couple to
$\bar{\bds\phi}$ with strength $(N-1)\lbare/N\to \lbare$; when $N=1$
only the longitudinal mode is there.

As $L\to\infty$, $\om^2_k \to k^2 +M^2$ and $M$ is exactly the
physical mass gap. Hence it must be $\Lambda-$independent.  At finite
$L$ we cannot use this request to determine $\mbare$ and $\lbare$,
since, unlike $M$, they cannot depend on the size $L$. At infinite volume
we obtain
\begin{equation}\label{M}
	M^2 =\mbare+3\lbare[\bar\phi^2+I_D(M^2,\Lambda)] \;,\quad
	I_D(z,\Lambda) \equiv \int_{k^2\le\Lambda^2} 
	\dfrac{d^Dk}{(2\pi)^D} \dfrac1{2\sqrt{k^2+z}}
\end{equation}
When $\bar\phi=0$ this equation fixes the bare mass to be
\begin{equation}\label{m2ren}
	\mbare= m^2 -3\lbare I_D(m^2,\Lambda)
\end{equation}
where $m=M(\bar\phi=0)$ may be identified with the equilibrium
physical mass of the scalar particles of the infinite--volume Fock
space without symmetry breaking (see below). Now, the coupling
constant renormalization follows from the equalities
\begin{equation}\label{rengap}
\begin{split}
	M^2&=m^2+3\lbare[\bar\phi^2+I_D(M^2,\Lambda)-I_D(m^2,\Lambda)]\\
	&=m^2+3\l\,\bar\phi^2+3\l \left[ I_D(M^2,\Lambda)-
	I_D(m^2,\Lambda) \right]_{\text{finite}} 
\end{split}
\end{equation}
and reads when $D=3$
\begin{equation}\label{lren}
	\frac\l\lbare = 1-\frac{3\l}{8\pi^2} \log\frac{2\Lambda}{m\sqrt{e}} 
\end{equation}
that is the standard result of the one--loop renormalization group
\cite{zj}.  When $D=1$, that is a $1+1-$dimensional quantum field
theory, $I_D(M^2,\Lambda)-I_D(m^2,\Lambda)$ is already finite and the
dimensionfull coupling constant is not renormalized, $\lbare=\l$.

The Landau pole in $\lbare$ prevents the actual UV limit
$\Lambda\to\infty$. Nonetheless, neglecting all inverse powers of the
UV cutoff when $D=3$, it is possible to rewrite the gap equation
(\ref{rengap}) as
\begin{equation}\label{nice2}
	\frac{M^2}{\hat\l(M)} = \frac{m^2}{\hat\l(m)} + 3\,\bar\phi^2
\end{equation}
in terms of the one--loop running coupling constant
\begin{equation*}
	\hat\l(\mu) = \l \left[ 1 - \frac{3\l}{8\pi^2} 
	\log\frac{\mu}m \right]^{-1}
\end{equation*}
It is quite clear that the HF states for which the renormalization
just defined is sufficient are all those that are gaussian--dominated
in the ultraviolet, so that we have [compare to eq. (\ref{gauss})]
\begin{equation}\label{largek}
	\vev{\varphi_{\pm k}^2} \sim \dfrac1{2\om_k} 
	\;,\quad k^2 \sim \Lambda^2\;,\; \Lambda \to \infty
\end{equation}
If this property holds at a certain time, then it should hold at all
times, since the Schroedinger equations (\ref{Schroedinger}) are
indeed dominated by the quadratic term for large $\om_k$ and
$\om^2_k\sim k^2+\text{const}+O(k^{-1})$ as evident from
eq. (\ref{renomvev}). Thus this class of states is indeed closed under
time evolution and the parameterizations (\ref{m2ren}) and (\ref{lren})
make our tdHF approximation ultraviolet finite. Notice that the
requirement (\ref{largek}) effectively always imposes a gap equation
similar to eq. (\ref{gap}) in the deep ultraviolet.

Another simple check of the self--consistency of our approach,
including the change in selected places from $\lbare$ to $\l$, as
discussed above, follows from the energy calculation for the gaussian
states with $\vev{\phi(x)}=\bar\phi$ introduced above. Using
eq. (\ref{energy}) and the standard replacement of sums by integrals
in the infinite volume limit, we find
\begin{equation*}
	\E(\bar\phi) = \lim_{L\to\infty} \frac{E}{L ^D} = 
	\tfrac12\bar\phi^2(M^2-\l\bar\phi^2)+\tfrac12\int_{k^2\le\Lambda^2} 
	\dfrac{d^Dk}{(2\pi)^D} \,\sqrt{k^2+M^2} -\tfrac34 \lbare 
	\left[\bar\phi^2+I_D(M^2,\Lambda)\right]^2
\end{equation*}
where $M=M(\bar\phi)$ depends on $\bar\phi$ through the gap equation
(\ref{rengap}). The explicit calculation of the integrals involved
shows that the energy density difference $\E(\bar\phi)-\E(0)$ [which
for unbroken symmetry is nothing but the effective potential
$V_{\text{eff}}(\bar\phi)$], is indeed finite in the limit
$\Lambda\to\infty$, as required by a correct renormalization
scheme. Notice that the finiteness of the energy density difference
can be shown also by a simpler and more elegant argument, as
presented below in section \ref{ooed}. This check would fail instead when
$D=3$ if only the bare coupling constant $\lbare$ would appear in the
last formula.

The tdHF approximation derived above represents a huge simplification
with respect to the original problem, but its exact solution still
poses itself as a considerable challenge. As a matter of fact, a
numerical approach is perfectly possible within the capabilities of
modern computers, provided the number of equations
(\ref{Schroedinger}) is kept in the range of few thousands. As will
become clear later on, even this numerical workout will turn out not
to be really necessary in the form just alluded to, at least for the
purposes of this paper.

\subsection{On symmetry breaking}

Quite obviously, in a finite volume and with a UV cutoff there cannot
be any symmetry breaking, since the ground state is necessarily unique
and symmetric when the number of degrees of freedom is finite
\cite{gj}. However, we may handily envisage the situation which would
imply symmetry breaking when the volume diverges.

Let us first consider the case that we would call of unbroken
symmetry. In this case the HF ground state is very close to the member
with $\bar\phi=0$ of the family of gaussian states introduced
before. The difference is entirely due to the quartic terms in $H_k$.
This correction vanish when $L\to\infty$, since all wavefunctions
$\psi_k$ have $L-$independent widths, so that one directly obtains the
symmetric vacuum state with all the right properties of the vacuum
(translation invariance, uniqueness, etc.)  upon which a standard scalar
massive particle Fock space can be based. The HF approximation then
turns out to be equivalent to the resummation of all ``cactus
diagrams'' for the particle self--energy \cite{cactus}. In a finite
volume, the crucial property of this symmetric vacuum is that all
frequencies $\om_k^2$ are strictly positive. The generalization to
non--equilibrium initial states with $\bar\phi\neq0$ is rather
trivial: it amounts to a shift by $L^{D/2}\bar\phi$ on
$\psi_0(\varphi_0)$. In the limit $L\to\infty$ we should express
$\psi_0$ as a function of $\xi=L^{-D/2}\varphi_0$ so that,
$|\psi_0(\xi)|^2\to \delta(\xi-\bar\phi)$, while all other
wavefunctions $\psi_k$ will reconstruct the gaussian wavefunctional
corresponding to the vacuum $\ket{0,M}$ of a free massive scalar
theory whose mass $M=M(\phi)$ solves the gap equation
(\ref{rengap}).  The absence of $\psi_0$ in $\ket{0,M}$ is
irrelevant in the infinite volume limit, since
$\vev{\varphi_0^2}=L^D\bar\phi^2 +$ terms of order $L^0$. The effective
potential $V_{\text{eff}}(\bar\phi)= \E(\bar\phi)-\E(0)$, where
$\E(\bar\phi)$ is the lowest energy density at fixed $\bar\phi$ and
infinite volume, is manifestly a convex function with a unique minimum
in $\bar\phi=0$.

Now let us consider a different situation in which one or more of the
$\om_k^2$ are negative. Quite evidently, this might happen only for
$k$ small enough, due to the $k^2$ in the gap equation [thus
eq. (\ref{largek}) remains valid and the ultraviolet renormalization
is the same as for unbroken symmetry]. Actually we assume here that
only $\om_0^2<0$, postponing the general analysis.  Now the quartic
term in $H_0$ cannot be neglected as $L\to\infty$, since in the ground
state $\psi_0$ is symmetrically concentrated around the two minima of
the potential $\frac12 \om_0^2 \varphi_0^2 +\frac\l{4L^D}\varphi_0^4$,
that is $\varphi_0=\pm(-\om_0^2L^D/\l)^{1/2}$.  If we scale
$\varphi_0$ as $\varphi_0=L^{D/2}\xi$ then $H_0$ becomes
\begin{equation}\label{Hscaled}
  H_0 = -\dfrac1{2L^D}\pdif{^2}{\xi^2}+\dfrac{L^D}2 \left(
	\om_0^2\, \xi^2 +\dfrac\l2\xi^4 \right)
\end{equation}
so that the larger $L$ grows the narrower $\psi_0(\xi)$ becomes around
the two minima $\xi=\pm(-\om_0^2/\l)^{1/2}$. In particular
$\vev{\xi^2}\to -\om_0^2/\l$ when $L\to\infty$ and
$\vev{\varphi_0^2}\simeq L^D\vev{\xi^2}$. Moreover, the energy gap
between the ground state of $H_0$ and its first, odd excited state as
well as difference between the relative probability distributions for $\xi$
vanish exponentially fast in the volume $L^D$. 
  
Since by hypothesis all $\om_k^2$ with $k\neq0$ are strictly positive,
the ground state $\psi_k$ with $k\neq0$ are asymptotically gaussian when
$L\to\infty$ and the relations (\ref{omvev2}) tend to the form 
\begin{equation*}
\begin{split}
	\om_k^2 &= k^2+ M^2  \equiv  k^2 +m^2 \\
	M^2 &= -2\om_0^2 = \mbare + 3\lbare (L^{-D}\vev{\varphi_0^2}
	+ \Sigma_0) = \mbare + 3\lbare\om_0^2+ 3\lbare I_D(m^2,\Lambda)] 
\end{split}
\end{equation*}  
This implies the identification $\om_0^2=-m^2/2$ and the bare mass
parameterization
\begin{equation}\label{brokenm}
	\mbare = \left(1-\tfrac32 \lbare/\l\right)m^2 -3\lbare I_D(m^2,\Lambda)
\end{equation}
characteristic of a negative $\om_0^2$ [compare to eq. (\ref{m2ren})],
with $m$ the physical equilibrium mass of the scalar particle, as in
the unbroken symmetry case. The coupling constant renormalization is
the same as in eq. (\ref{lren}) as may be verified by generalizing to
the minimum energy states with given field expectation value
$\bar\phi$; this minimum energy is nothing but the HF effective potential
$V_{\text{eff}}^{\text{HF}}(\bar\phi)$, that the effective potential
in this non--gaussian HF approximation; of course, since $\psi_0$ is no longer
asymptotically gaussian, we cannot simply shift it by $L^{D/2}\bar\phi$
but, due to the concentration of $\psi_0$ on classical minima as
$L\to\infty$, one readily finds that $V_{\text{eff}}(\bar\phi)$ is the
convex envelope of the classical potential, that is its Maxwell
construction. Hence we find
\begin{equation*}
	\vev{\varphi_0^2} \underset{L\to\infty}\sim
	\begin{cases}
	-L^D\om_0^2/\l \;,\; & \l\bar\phi^2\le-\om_0^2 \\
	L^D\bar\phi^2  \;,\; & \l\bar\phi^2>-\om_0^2
	\end{cases}
\end{equation*}
and the gap equation for the $\bar\phi-$dependent mass $M$ can be
written, in terms of the step function $\Theta$ and the extremal
ground state field expectation value $v=m/\sqrt{2\l}$,
\begin{equation}\label{gapbroken}
	M^2 = m^2 + 3\lbare(\bar\phi^2-v^2) \,
	\Theta(\bar\phi^2-v^2) +3\lbare \left[ 
	I_D(M^2,\Lambda)-I_D(m^2,\Lambda) \right] 
\end{equation}
We see that the specific bare mass parameterization (\ref{brokenm})
guarantees the non--renormalization of the tree--level relation
$v^2=m^2/2\l$ ensuing from the typical symmetry breaking classical
potential $V(\phi)=\frac14\l(\phi^2-v^2)^2$. With the same finite part
prescription as in eq. (\ref{rengap}), the gap equation
(\ref{gapbroken}) leads to the standard coupling constant
renormalization (\ref{lren}) when $D=3$.

In terms of the probability distributions $|\psi_0(\xi)|^2$ for the
scaled amplitude $\xi=L^{-D/2}\varphi_0$, the Maxwell construction
corresponds to the limiting form
\begin{equation}\label{limitform}
	|\psi_0(\xi)|^2 \underset{L\to\infty}\sim
	\begin{cases}
	\tfrac12(1+\bar\phi/v)\delta(\xi-v)+ \tfrac12(1-\bar\phi/v)\,
	\delta(\xi+v) \;,\; &\bar\phi^2 \le v^2 \\
	\delta(\xi-\bar\phi) \;,\;  &\bar\phi^2 > v^2 
	\end{cases}
\end{equation}
On the other hand, if $\om^2_0$ is indeed the
only negative squared frequency, the $k\ne0$ part of this minimum
energy state with arbitrary $\bar\phi=\vev{\phi(x)}$ is better and
better approximated as $\L\to\infty$ by the same gaussian state
$\ket{0,M}$ of the unbroken symmetry state. Only the effective mass
$M$ has a different dependence $M(\bar\phi)$, as given by the
gap equation (\ref{gapbroken}) proper of broken symmetry.

At infinite volume we may write
\begin{equation*}
	\vev{\varphi_k^2} = C(\bar\phi)\,\delta^{(D)}(k)+
	\frac1{2\sqrt{k^2+M^2}}
\end{equation*}
where $C(\bar\phi)=\bar\phi^2$ in case of unbroken symmetry (that is
$\om^2_0>0$), while $C(\bar\phi)=\text{max}(v^2,\bar\phi^2)$
when $\om^2_0<0$.  This corresponds to the field correlation in space
\begin{equation*}
	\vev{\phi(x)\phi(y)}=\int\frac{d^Dk}{(2\pi)^D}
	\vev{\varphi_k^2} e^{ik\cdot(x-y)}= C(\bar\phi) +\Delta_D(x-y,M)
\end{equation*}
where $\Delta_D(x-y,M)$ is the massive free field equal--time two
points function in $D$ space dimensions, with self--consistent mass
$M$. The requirement of clustering
\begin{equation*}
	\vev{\phi(x)\phi(y)} \to \vev{\phi(x)}^2 = v^2 
\end{equation*}
contradicts the infinite volume limit of 
\begin{equation*}
	\vev{\phi(x)}=L^{-D/2}\sum_k \vev{\phi_k}\,e^{ik\cdot x} =
	\vev{\varphi_0} = \bar\phi
\end{equation*}
except at the two extremal points $\bar\phi=\pm v$. In fact we know
that the $L\to\infty$ limit of the finite volume states with
$\bar\phi^2<v^2$ violate clustering, because the two peaks of
$\psi_0(\xi)$ have vanishing overlap in the limit and the first
excited state becomes degenerate with the vacuum: this implies that
the relative Hilbert space splits into two orthogonal Fock sectors
each exhibiting symmetry breaking, $\vev{\phi(x)}=\pm v$, and
corresponding to the two independent equal weight linear combinations
of the two degenerate vacuum states. The true vacuum is either one of
these symmetry broken states. Since the two Fock sectors are not only
orthogonal, but also superselected (no local observable interpolates
between them), linear combinations of any pair of vectors from the two
sectors are not distinguishable from mixtures of states and clustering
cannot hold in non--pure phases. It is perhaps worth noticing also
that the Maxwell construction for the effective potential, in the
infinite volume limit, is just a straightforward manifestation of this
fact and holds true, as such, beyond the HF approximation.

To further clarify this point and in view of subsequent
applications, let us consider the probability distribution for the
smeared field $\phi_f=\int d^Dx\,\phi(x)f(x)$, where
\begin{equation*}
	f(x)=f(-x)=\frac1{L^D}\sum_k f_k \,e^{ik\cdot x} 
	\underset{L\to\infty}\sim \,\int \dfrac{d^Dk}{(2\pi)^D}
	\, \tilde f(k) \,e^{ik\cdot x} 
\end{equation*}
is a smooth real function with $\int d^Dx\,f(x)=1$ ({\em
i.e.} $f_0=1$) localized around the origin (which is good as any other
point owing to translation invariance). Neglecting in the infinite
volume limit the quartic corrections for all modes with $k\neq 0$, so
that the corresponding  ground state wavefunctions are
asymptotically gaussian, this probability distribution evaluates to
\begin{equation*}
	\text{Pr}(u\!<\!\phi_f\!<\!u+du)= \frac{du}{(2\pi\Sigma_f)^{1/2}} 
	\intf d\xi\, |\psi_0(\xi)|^2
	\exp\left\{\frac{-(u-\xi)^2}{2\Sigma_f} \right\}
\end{equation*}
where
\begin{equation*}
	\Sigma_f = \sum_{k\neq0}\vev{\varphi_k^2}\,f_k^2 
	\;\underset{L\to\infty}\sim \;\int \dfrac{d^Dk}{(2\pi)^D} 
	\,\dfrac{\tilde f(k)^2}{2\sqrt{k^2+m^2}}
\end{equation*}
In the unbroken symmetry case we have
$|\psi_0(\xi)|^2\sim\delta(\xi-\bar\phi)$ as $L\to\infty$, while
the limiting form (\ref{limitform}) holds for broken symmetry.
Thus we obtain
\begin{equation*}
	\text{Pr}(u\!<\!\phi_f\!<\!u+du) =p_f(u-\bar\phi)\,du \;,\quad 
	p_f(u) \equiv \left(2\pi\Sigma_f\right)^{-1/2}
	\exp\left(\frac{-u^2}{2\Sigma_f} \right)
\end{equation*}
for unbroken symmetry and 
\begin{equation*}
	\text{Pr}(u\!<\!\phi_f\!<\!u+du) =\begin{cases}
	\tfrac12(1+\bar\phi/v)\,p_f(u-v)\,du+\tfrac12(1-\bar\phi/v)
	\,p_f(u+v)\,du \;, &\bar\phi^2 \le v^2 \\
	p_f(u-\bar\phi)\,du\;, &\bar\phi^2 > v^2 
	\end{cases}
\end{equation*}
for broken symmetry. Notice that the momentum integration in the
expression for $\Sigma_f$ needs no longer an ultraviolet cutoff; of
course in the limit of delta--like test function $f(x)$, $\Sigma_f$
diverges and $p_f(u)$ flattens down to zero. The important observation
is that $\text{Pr}(u\!<\!\phi_f\!<\!u+du)$ has always a single peak
centered in $u=\bar\phi$ for unbroken symmetry, while for broken symmetry it
shows two peaks for $\bar\phi^2 \le v^2$ and  $\Sigma_f$ small enough.
For instance, if $\bar\phi=0$, then there are two peaks for 
$\Sigma_f<v^2$ [implying that $\tilde f(k)$ has a significant
support only up to wavevector $k$ of order $v$, when $D=3$, or
$m\exp(\text{const }v^2)$ when $D=1$].

To end the discussion on symmetry breaking, we may now verify the
validity of the assumption that only $\om_0^2$ is negative. In fact,
to any squared frequency $\om_k^2$ (with $k\neq0$) that stays strictly
negative as $L\to\infty$ there corresponds a wavefunction $\psi_k$
that concentrates on $\varphi_k^2+\varphi_{-k}^2=-\om_k^2L^D/\l$ ;
then eqs. (\ref{omvev2}) implies $-2\om_k^2 = k^2 + m^2$ for such
frequencies, while $\om_k^2 = k^2 +m^2$ for all frequencies with
positive squares; if there is a macroscopic number of negative
$\om_k^2$ (that is a number of order $L^D$), then the expression for
$\om_0^2$ in eq. (\ref{omvev2}) will contain a positive term of order
$L^D$ in the r.h.s., clearly incompatible with the requirements that
$\om_0^2<0$ and $\mbare$ be independent of $L$; if the number of
negative $\om_k^2$ is not macroscopic, then the largest wavevector
with a negative squared frequency tends to zero as $L\to\infty$ (the
negative $\om_k^2$ clearly pile in the infrared) and the situation is
equivalent, if not identical, to that discussed above with only
$\om_0^2<0$.

\subsection{Out--of--equilibrium dynamics}\label{ooed}

We considered above the lowest energy states with a predefinite
uniform field expectation value, $\vev{\phi(x)}=\bar\phi$, and
established how they drastically simplify in the infinite volume
limit.  For generic $\bar\phi$ these states are not stationary and
will evolve in time. By hypothesis $\psi_k$ is the ground state
eigenfunction of $H_k$ when $k>0$, and therefore $|\psi_k|^2$ would be
stationary for constant $\om_k$, but $\psi_0$ is not an eigenfunction
of $H_0$ unless $\bar\phi=0$. As soon as $|\psi_k|^2$ starts changing,
$\vev{\varphi_0^2}$ changes and so do all frequencies $\om_k$ which
are coupled to it by the eqs. (\ref{omvev2}). Thus the change
propagates to all wavefunctions.  The difficult task of studying this
dynamics can be simplified with the following scheme, that we might
call {\em gaussian approximation}. We first describe it and discuss
its validity later on.

Let us assume the usual gaussian form for the initial state [see
eq. (\ref{gauss}) and the discussion following it]. We know that it is
a good approximation to the lowest energy state with given
$\vev{\varphi_0}$ for unbroken symmetry, while it fails to be so for
broken symmetry, only as far as $\psi_0$ is concerned, unless
$\bar\phi^2 \geq v^2$. At any rate this is an acceptable initial state: the
question is about its time evolution. Suppose we adopt the harmonic
approximation for all $H_k$ with $k>0$ by dropping the quartic term.
This approximation will turn out to be valid only if the
width of $\psi_k$ do not grow up to the order $L^D$ (by symmetry the
center will stay in the origin). In practice we are now dealing with a
collection of harmonic oscillators with time--dependent frequencies and
the treatment is quite elementary: consider the simplest example of one quantum
degree of freedom described by the gaussian wavefunction
\begin{equation*}
	\psi(q,t)=\frac1{(2\pi\s^2)^{1/4}}
	\exp\left[-\frac12\left(\frac1{2\s^2}-i\frac{s}\s\right)q^2\right]
\end{equation*}
where $s$ and $\s$ are time--dependent. If the dynamics is determined
by the time--dependent harmonic hamiltonian
$\frac12[-\partial^2_q+\om(t)^2\,q^2]$, then the Schroedinger equation is
solved exactly provided that $s$ and $\s$ satisfy the classical
Hamilton equations
\begin{equation*}
	\dot\s = s \;,\quad \dot s = - \om^2\s + \frac1{4 \s^3}
\end{equation*}
It is not difficult to trace the ``centrifugal'' force $(4\s)^{-3}$
which prevents the vanishing of $\s$ to Heisenberg uncertainty
principle \cite{chkm,hkmp}.

The extension to our case with many degrees of freedom is
straightforward and we find the following system of equations
\begin{equation}\label{sk}
	i\pdif{}t \psi_0 = H_0\psi_0 \;,\quad
	\der{^2 \s_k}{t^2}= - \om_k^2\,\s_k + \frac1{4 \s_k^3} \;,\; k>0
\end{equation}
coupled in a mean--field way by the relations (\ref{omvev2}), which
now read
\begin{equation}\label{momvev2}
\begin{split}
	\om_k^2 &= k^2+ M^2 - 6\l\, L^{-D}\s_k^2 \;,\quad k>0 \\ 
	M^2 &= \mbare +3\lbare \left(L^{-D}\vev{\varphi_0^2} + \Sigma_0\right)
	\;,\quad \Sigma_0= \frac1{L^D} \sum_{k\ne 0}\s_k^2 
\end{split}
\end{equation}
This stage of a truly quantum zero--mode and classical modes with
$k>0$ does not appear fully consistent, since for large volumes some type of
classical or gaussian approximation should be considered for
$\varphi_0$ too. We may proceed in two (soon to be proven equivalent)
ways:
\begin{enumerate}
\item 
We shift $\varphi_0=L^{D/2}\bar\phi+\eta_0$ and then
deal with the quantum mode $\eta_0$ in the gaussian approximation,
taking into account that we must have $\vev{\eta_0}=0$ at all times.
This is most easily accomplished in the Heisenberg picture rather than
in the Schroedinger one adopted above. In any case we find that
the quantum dynamics of $\varphi_0$ is equivalent to the classical
dynamics of $\bar\phi$ and $\s_0\equiv\vev{\eta_0^2}^{1/2}$ described
by the ordinary differential equations
\begin{equation}\label{classical}
	\der{^2 \bar\phi}{t^2} = -\om_0^2\, \bar\phi -\l\, \bar\phi^3
	\;,\quad \der{^2 \s_0}{t^2}= - \om_0^2\,\s_0 + \frac1{4 \s_0^3}
\end{equation}
where $\om_0^2=M^2-3\l\,L^{-D}\vev{\varphi_0^2}$ and 
$\vev{\varphi_0^2} = L^D\bar\phi^2+ \s_0^2$.

\item 
We rescale $\varphi_0=L^{D/2}\xi$ right away, so that $H_0$ takes the
form of eq. (\ref{Hscaled}). Then $L\to\infty$ is the classical limit
such that $\psi_0(\xi)$ concentrates on $\xi=\bar\phi$ which evolves
according to the first of the classical equations in
(\ref{classical}). Since now there is no width associated to the
zero--mode, $\bar\phi$ is coupled only to the widths $\s_k$ with $k\neq 0$
by $\om_0^2=M^2-3\l\bar\phi^2$, while $M^2=\mbare
+3\lbare(\bar\phi^2+\Sigma_0)$.
\end{enumerate}

It is quite evident that these two approaches are completely
equivalent in the infinite volume limit, and both are good
approximation to the original tdHF Schroedinger equations, at least
provided that $\s_0^2$ stays  such that
$L^{-D}\s_0^2$ vanishes in the limit for any time. In this case we
have the evolution equations
\begin{equation}\label{Emotion}
	\der{^2 \bar\phi}{t^2} = (2\l\,\bar\phi^2 -M^2)\,\bar\phi \;,\quad
	\der{^2\s_k}{t^2} = - (k^2+M^2)\,\s_k + \frac1{4 \s_k^3} 	
\end{equation}
mean--field coupled by the $L\to\infty$ limit of eqs. (\ref{momvev2}),
namely
\begin{equation}\label{unbtdgap}
	M^2=m^2+3\lbare\left[\bar\phi^2 + \Sigma -I_D(m^2,\Lambda)\right]
\end{equation}
for unbroken symmetry [that is $\mbare$ as in eq. (\ref{m2ren})] or
\begin{equation}\label{btdgap}
	M^2=m^2+3\lbare\left[\bar\phi^2 -v^2 + 
	\Sigma -I_D(m^2,\Lambda)\right] \;,\quad m^2=2\l v^2
\end{equation}
for broken symmetry [that is $\mbare$ as in
eq. (\ref{brokenm})]. In any case we define
\begin{equation*}
	 \Sigma =\frac1{L^D} \sum_k\s_k^2 \;\underset{L\to\infty}\sim\;
	\int_{k^2\le\Lambda^2} \dfrac{d^Dk}{(2\pi)^D}\,\s_k^2
\end{equation*}
as the sum, or integral, over all microscopic gaussian widths
[N.B.:this definition differs from that given before in
eq. (\ref{Sigma}) by the classical term $\bar\phi^2$]. Remarkably, the
equations of motion (\ref{Emotion}) are completely independent of the
ultraviolet cut--off and this is a direct consequence of the
substitution (\ref{H_k'}). Had we kept the bare coupling constant
everywhere in the expression (\ref{subst}), we would now have $\lbare$
also in front of the $\bar\phi^3$ in the r.h.s. of the first of the two
equations (\ref{Emotion}) [cfr., for instance, ref. \cite{devega2}].

The conserved HF energy (density) corresponding to these equations of
motion reads
\begin{equation}\label{EHF}
\begin{split}
	\E &= \T +\V \;,\quad \T = \frac12(\dot{\bar{\phi}})^2 +
	\frac1{2L^D}\sum_k \dot\s_k^2  \\ \V &= \frac1{2L^D}\sum_k\left(
	 k^2\,\s_k^2 + \frac1{4\s^2_k}\right) +\tfrac12\mbare(\bar\phi^2+
	\Sigma) + \tfrac34\lbare (\bar\phi^2+\Sigma)^2 -\tfrac12\l\bar\phi^4
\end{split}
\end{equation}
Up to additive constants and terms vanishing in the infinite volume
limit, this expression agrees with the general HF energy of
eq. (\ref{menergy}) for gaussian wavefunctions.  It holds both for
unbroken and broken symmetry, the only difference being in the
parameterization of the bare mass in terms of UV cutoff and physical
mass, eqs. (\ref{m2ren}) and (\ref{brokenm}). The similarity to the
energy functional of the large $N$ approach is evident; the only
difference, apart from the obvious fact that $\bar{\phi}$ is a single
scalar rather than a $O(n)$ vector, is in the mean--field coupling
$\s_k$--$\bar\phi$ and $\s_k$--$\Sigma$, due to different coupling
strength of transverse and longitudinal modes
(cfr. ref. \cite{finvN}).

This difference between the HF approach for discrete symmetry ({\em
i.e} $N=1$) and the large $N$ method for the continuous
$O(N)$-symmetry is not very relevant if the symmetry is unbroken [it
does imply however a significantly slower dissipation to the modes of
the background energy density].  On the other hand it has a drastic
consequence on the equilibrium properties and on the
out--of--equilibrium dynamics in case of broken symmetry (see below),
since massless Goldstone bosons appear in the large $N$ approach,
while the HF treatment of the discrete symmetry case must exhibits a
mass also in the broken symmetry phase.

The analysis of physically viable initial conditions proceeds exactly
as in the large $N$ approach \cite{finvN} and will not be repeated
here, except for an important observation in case of broken
symmetry. The formal energy minimization w.r.t. $\s_k$ at fixed
$\bar\phi$ leads again to eqs. 
\begin{equation}\label{inis}
	\dot \s_k=0 \;,\quad \s_k^2=\frac1{2\sqrt{k^2+M^2}}
\end{equation}
and again these are acceptable initial conditions only if the gap
equation that follows from eq. (\ref{btdgap}) in the $L\to\infty$
limit, namely
\begin{equation}\label{bgap}
	M^2 = m^2 + 3\lbare \left[\bar\phi^2-v^2 +
	I_D(M^2,\Lambda)-I_D(m^2,\Lambda) \right] 
\end{equation}
admits a nonnegative, physical solution for $M^2$. Notice that there
is no step function in eq. (\ref{bgap}), unlike the static case of
eq. (\ref{gapbroken}), because $\s_0^2$ was assumed to be microscopic,
so that the infinite volume $\s_k^2$ has no delta--like singularity in
$k=0$. Hence $M=m$ solves eq. (\ref{bgap}) only at the extremal points
$\bar\phi=\pm v$, while it was the solution of the static gap equation
(\ref{gapbroken}) throughout the Maxwell region $-v\le\bar\phi\le v$.
The important observation is that eq. (\ref{bgap}) admits a positive
solution for $M^2$ also within the Maxwell region. In fact it can be
written, neglecting as usual the inverse--power corrections in the UV
cutoff
\begin{equation}\label{bnice}
	\frac{M^2}{\hat\l(M)} = \frac{m^2}\l + 3\,(\bar\phi^2 -v^2) =
	3\,\bar\phi^2 -v^2
\end{equation}
and there exists indeed a positive solution $M^2$ smoothly connected
to the ground state, $\bar\phi^2=v^2$ and $M^2=m^2$, whenever
$\bar\phi^2\ge v^2/3$. The two intervals $v^2\ge\bar\phi^2\ge v^2/3$
correspond indeed to the metastability regions, while $\bar\phi^2<
v^2/3$ is the spinodal region, associated to a classical potential
proportional to $(\bar\phi^2 -v^2)^2$. This is another effect of the
different coupling of transverse and longitudinal modes: in the large
$N$ approach there are no metastability regions and the spinodal
region coincides with the Maxwell one. As in the large $N$ approach in
the spinodal interval there is no energy minimization possible, at
fixed background and for microscopic widths, so that a modified form
of the gap equation
\begin{equation}\label{newgap}
	M^2 = m^2 + 3\lbare \left[\bar{\phi}^2-v^2 + \frac1{L^D} 
	\sum_{k^2<|M^2|}\s_k^2 + \frac1{L^D}\sum_{k^2>|M^2|}
	\frac1{2\sqrt{k^2-|M^2|}} -I_D(0,\Lambda) \right] 
\end{equation}
should be applied to determine ultraviolet--finite initial conditions.

The main question now is: how will the gaussian widths $\s_k$ grow
with time, and in particular how will $\s_0$ grow in case of method 1
above, when we start from initial conditions where all widths are
microscopic? For the gaussian approximation to remain valid through
time, all $\s_k$, and in particular $\s_0$, must at least not become
macroscopic.  In fact we have already positively answered this
question in the large $N$ approach \cite{finvN} and the HF equations
(\ref{Emotion}) do not differ so much to expect the contrary now. In
particular, if we consider the special initial condition
$\bar\phi=\dot{\bar\phi}=0$, the dynamics of the widths is identical
to that in the large $N$ approach, apart from the rescaling by a
factor of three of the coupling constant.

\begin{figure} %1
\epsfig{file=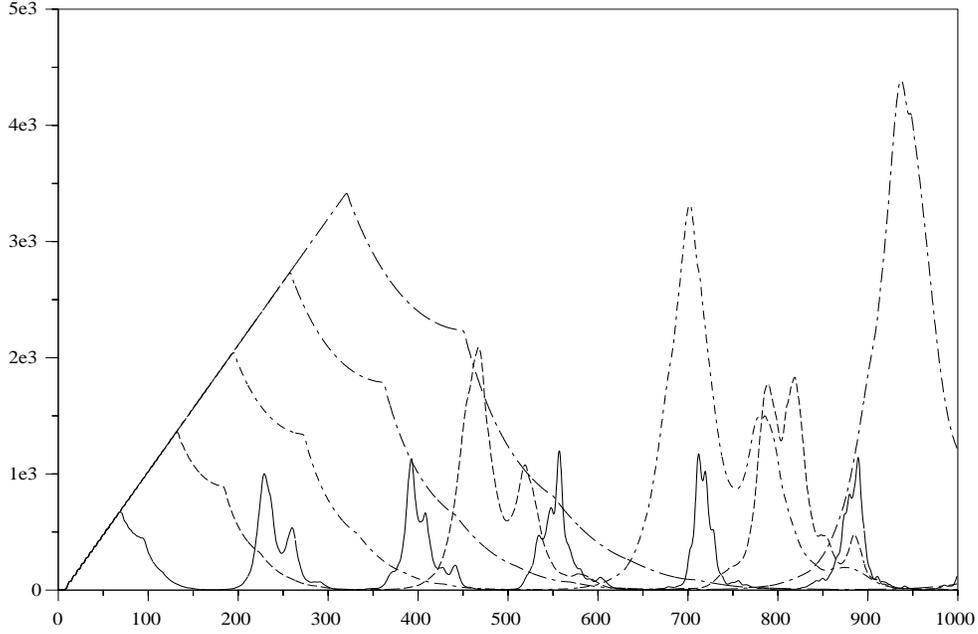,height=10cm,width=15cm}
\caption{\it Zero--mode amplitude evolution for 
different values of the size
$\frac{L}{2\pi}=20,40,60,80,100$, for $\lambda = 0.1$ and broken symmetry,
with $\bar\phi=0$. }\label{fig:m0}
\end{figure}

In fact, if we look at the time evolution of the zero--mode amplitude
$\s_0$ [see Fig. \ref{fig:m0}], we can see the presence of the time--scale
$\tau_L$ at which finite volume effects start to manifest. The time
scale $\tau_L$ turns out to be proportional to the linear size of the
box $L$ and its presence prevents $\s_0$ from growing to macroscopic
values. Thus our HF approximation confirms the large $N$ approach in
the following sense: even if one considers in the variational ansatz
the possibility of non--gaussian wavefunctionals, the time evolution
from gaussian and microscopic initial conditions is effectively
restricted for large volumes to non--macroscopic gaussians.

\begin{figure} %2
\epsfig{file=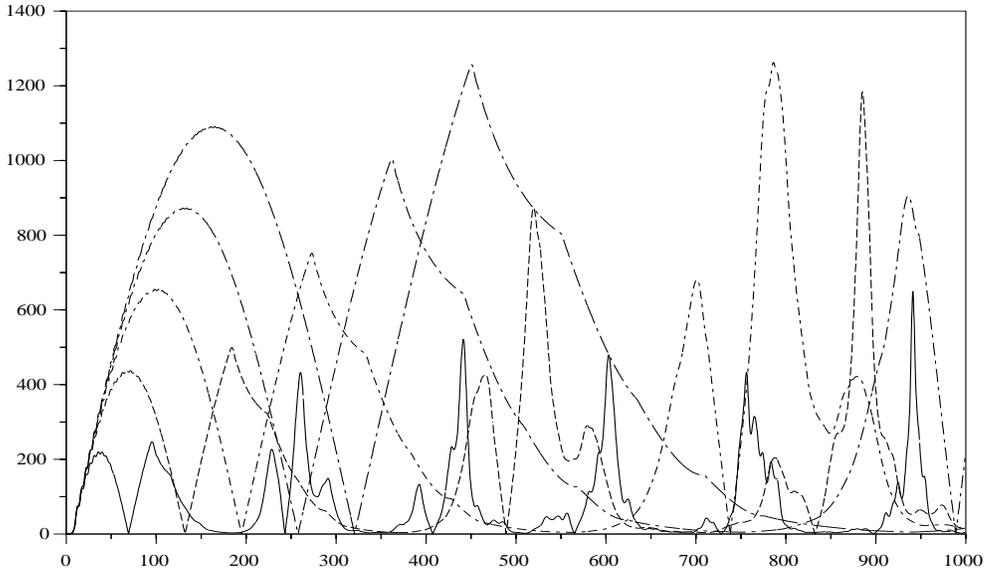,height=9cm,width=15cm}
\caption{\it Next--to--zero mode ($k=2\pi/L$) amplitude evolution for
different values of the size $L=20,40,60,80,100$, for $\lambda =
0.1$ and broken symmetry, with $\bar\phi=0$.}\label{fig:m1}
\end{figure}

Strictly speaking, however, this might well not be enough, since the
infrared fluctuations do grow beyond the microscopic size to become of
order $L$ [see Fig. \ref{fig:m1}, where the evolution of the mode with momentum
$k=2\pi/L$ is plotted]. Then the quartic term in the low$-k$
Hamiltonians $H_k$ is of order $L$ and therefore it is not negligible
by itself in the $L\to\infty$ limit, but only when compared to the
quadratic term, which {\em for a fixed $\om_k^2$ of order $1$} would
be of order $L^2$. But we know that, when $\bar\phi=0$, after the
spinodal time and before the $\tau_L$, the effective squared mass
$M^2$ oscillates around zero with amplitude decreasing as $t^{-1}$ and
a frequency fixed by the largest spinodal wavevector. In practice it
is ``zero on average'' and this reflect itself in the average linear
growth of the zero--mode fluctuations and, more generally, in the
average harmonic motion of the other widths with non--zero
wavevectors. In particular the modes with small wavevectors of order
$L^{-1}$ feel an average harmonic potential with $\om_k^2$ of order
$L^{-2}$. This completely compensate the amplitude of the mode itself,
so that the quadratic term in the low$-k$ Hamiltonians $H_k$ is of
order $L^0$, much smaller than the quartic term that was neglected
beforehand in the gaussians approximation. Clearly the approximation
itself no longer appears fully justified and a more delicate analysis
is required. We intend to return on this issue in a future work,
restricting ourselves in the next section to the gaussians approximation.

\section{Late--time evolution and dynamical Maxwell construction}\label{ep}

By definition, the gaussian approximation of the effective potential
$V_{\text{eff}}(\bar\phi)$ coincides with the infinite--volume limit
of the potential energy $\V(\bar\phi,\{\s_k\})$ of eq. (\ref{EHF})
when the widths are of the $\bar\phi-$dependent, energy--minimizing
form (\ref{inis}) with the gap equation for $M^2$ admitting a
nonnegative solution. As we have seen, this holds true in the unbroken
symmetry case for any value of the background $\bar\phi$, so that the
gaussian $V_{\text{eff}}$ is identical to the HF one, since all
wavefunctions $\psi_k$ are asymptotically gaussians as
$L\to\infty$. In the presence of symmetry breaking instead, this
agreement holds true only for $\bar\phi^2\ge v^2$; for
$v^2/3\le\bar\phi^2<v^2$ the gaussian $V_{\text{eff}}$ exists but is
larger than the HF potential $V_{\text{eff}}^{\text{HF}}$, which is
already flat. In fact, for any $\bar\phi^2\ge v^2/3$, we may write the
gaussian $V_{\text{eff}}$ as
\begin{equation*}
	V_{\text{eff}}(\bar\phi) = V_{\text{eff}}(-\bar\phi) =
	V_{\text{eff}}(v) + \int_v^{|\bar\phi|}\,du\,u[M(u)^2-2\l\,u^2] 
\end{equation*}
where $M(u)^2$ solves the gap equation (\ref{bnice}), namely
$M(u)^2=\hat\l(M(u))(3u^2-v^2)$. In each of the two disjoint regions
of definition this potential is smooth and convex, with unique minima
in $+v$ and $-v$, respectively. These appear therefore as regions of
metastability (states which are only locally stable in the presence of
a suitable uniform external source). The HF effective potential is
identical for $\bar\phi^2\ge v^2$, while it takes the constant value
$V_{\text{eff}}(v)$ throughout the internal region
$\bar\phi^2<v^2$. It is based on truly stable (not only metastable)
states. The gaussian $V_{\text{eff}}$ cannot be defined in the
spinodal region $\bar\phi^2<v^2/3$, where the gap equation does not
admit a nonnegative solution in the physical region far away from the
Landau pole.

Let us first compare this HF situation with that of large $N$
\cite{finvN}.  There the different coupling of the transverse modes,
three time smaller than the HF longitudinal coupling, has two main
consequences at the static level: the gap equation similar to
(\ref{bnice}) does not admit nonnegative solutions for
$\bar{\bds\phi}^2<v^2$, so that the spinodal region coincides with the
region in which the effective potential is flat, and the physical mass
vanishes. The out--of--equilibrium counterpart of this is the
dynamical Maxwell construction: when the initial conditions are such
that $\bar{\bds\phi}^2$ has a limit for $t\to\infty$, the set of all
possible asymptotic values exactly covers the flatness region (and the
effective mass vanishes in the limit). In practice this means that
$|\bar{\bds\phi}|$ is not the true dynamical order parameter, whose
large time limit coincides with $v$, the equilibrium field expectation
value in a pure phase. Rather, one should consider as order parameter
the renormalized local (squared) width
\begin{equation*}
	\lim_{N \to \infty} \frac{\vev{\bds{\phi}(x)
	\cdot\bds{\phi}(x)}_{\text{R}}}{N} = \bar{\bds\phi}^2 +
	\Sigma_{\text{R}} =  v^2 + \frac{M^2}\l
\end{equation*}
where the last equality follows from the definition itself of the
effective mass $M$ (see ref. \cite{finvN}). Since $M$ vanishes as
$t\to\infty$ when $\bar{\bds\phi}^2$ tends to a limit within the
flatness region, we find the renormalized local width tends to
the correct value $v$ which characterizes the broken symmetry phase,
that is the bottom of the classical potential. We may say that the 
spinodal region, perturbatively unstable, at the non--perturbative
level corresponds to metastable states, all reachable through the 
asymptotic time evolution with a vanishing effective mass.

\begin{figure} %3
\epsfig{file=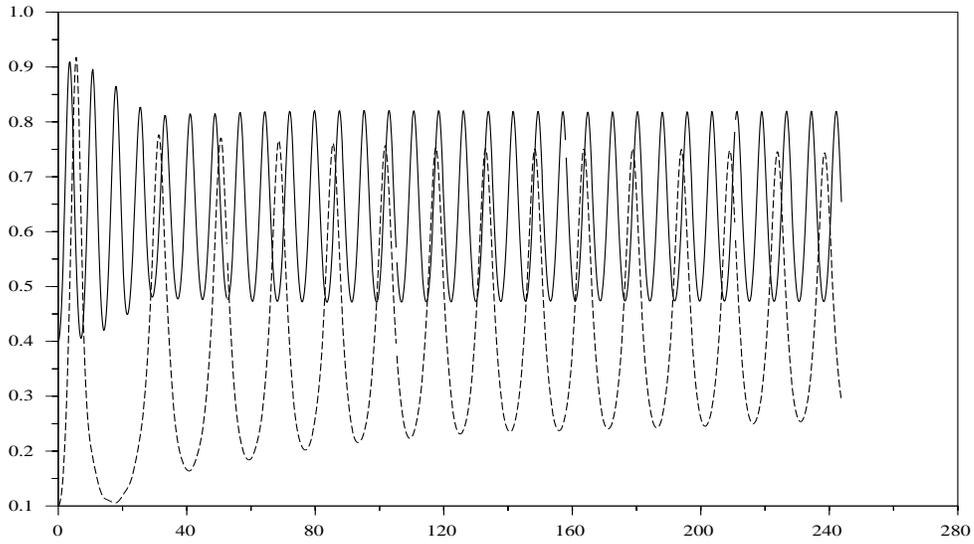,height=8.5cm,width=15cm}
\caption{\it Evolution of the background for two different initial
conditions within the spinodal interval, in the tdHF
approximation, for $\l=1$: $\bar\phi(t=0)=0.1$ (dotted line) and
$\bar\phi(t=0)=0.4$ (solid line). }\label{fig:max1}
\end{figure}

\begin{figure} %4
\epsfig{file=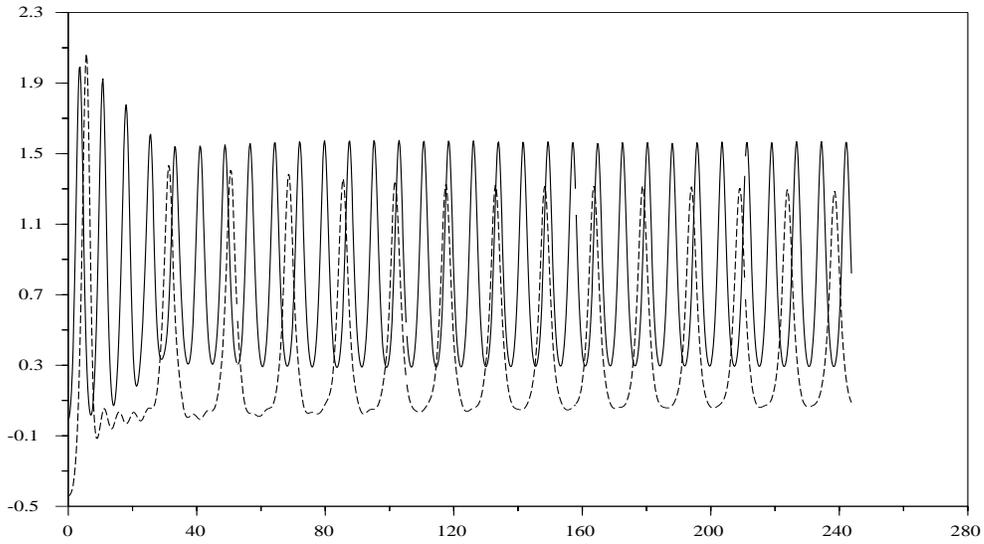,height=8.5cm,width=15cm}
\caption{\it Evolution of $M^2$ for the two initial
conditions of fig. \ref {fig:max1}. }\label{fig:max2}
\end{figure}

\begin{figure} %5
\epsfig{file=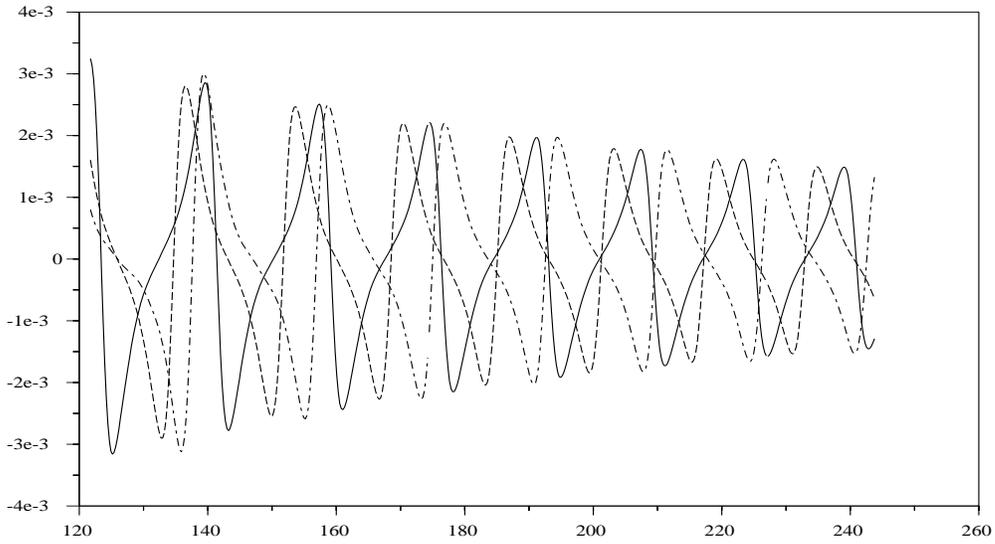,height=8.5cm,width=15cm}
\caption{\it The average force $f$, defined as
$\bar{f}=\int^Tf(t)dt/T$, plotted vs. $T$, for $\l=0.1$ and
$\bar\phi=10^{-2}$ (solid line), $\bar\phi=10^{-3}$ (dashed line) and
$\bar\phi=10^{-4}$ (dotted-dashed line).}\label{fig:meanforce}
\end{figure}

\begin{figure} %6
\epsfig{file=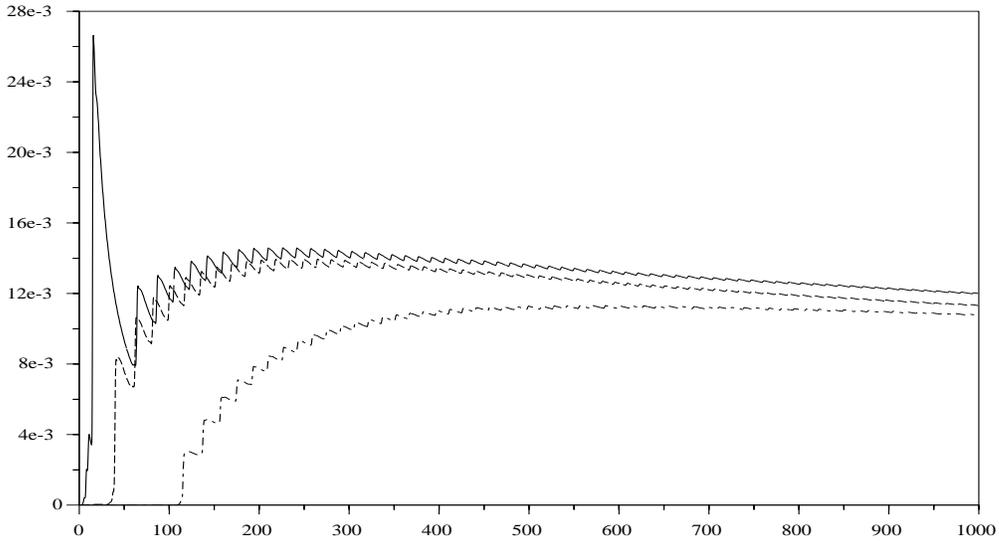,height=8.5cm,width=15cm}
\caption{\it The mean squared fluctuations of the force $f$, defined
as $\int^T(f(t)-\bar{f})^2dt/T$, plotted vs. $T$, for the three
initial conditions of fig. \ref{fig:meanforce}.}\label{fig:sq_flct}
\end{figure}

In the HF approximation, where at the static level the spinodal region
$\bar\phi^2<v^2/3$ is smaller than the flatness region
$\bar\phi^2<v^2$, the situation is rather different. Our numerical
solution shows that, $\bar\phi$ oscillates around a certain value
$\bar\phi _{\infty}$ with an amplitude that decreases very slowly. As
in large $N$, the asymptotic value $\bar\phi _{\infty}$ depends on the
initial value $\bar\phi (0)$. But, if the background $\bar{\phi}$
starts with zero velocity from a non--zero value inside the spinodal
interval, then it always leaves this region and eventually oscillates
around a point between the spinodal point $v/\sqrt{3}$ and the minimum
of the tree level potential $v$ (see Fig.s \ref{fig:max1} and
\ref{fig:max2}). In other words, if we start with a $\bar\phi$ in the
interval $[-v,v]$, except the origin, we end up with a $\bar\phi
_{\infty}$ in the restricted interval
$[-v,-v/\sqrt{3}]\cup[v/\sqrt{3},v]$. The spinodal region is
completely forbidden for the late time evolution of the mean field, as
is expected for an unstable region. We stress that we are dealing with
true fixed points of the asymptotic evolution since the force term on
the mean field [cfr. eq. (\ref{Emotion}), $f=(2\l\,\bar\phi^2
-M^2)\,\bar\phi$] does vanish in the limit. In facts its time average
$\bar{f}=\int^Tf(t)dt/T$ tends to zero as $T$ grows and its mean
squared fluctuations around $\bar{f}$ decreases towards zero, although
very slowly (see Fig.s \ref{fig:meanforce} and
\ref{fig:sq_flct}). Moreover, for $N=1$ the order parameter reads as
$t \to \infty$
\begin{equation}\label{aop}
	\vev{\phi(x)^2}_{\text{R}} = \bar{\phi}^2 +
	\Sigma_{\text{R}} = \frac{v^2}3 + 
	\frac{M^2}{3\l} \;,\quad 
	\Sigma_{\text{R}} = \frac{v^2 - \bar\phi ^2}{3}
\end{equation}
where the last equality is valid for the asymptotic values and follows
from the vanishing of the force term $f$. From the last formula we see
that when $\bar\phi=0$ at the beginning, and then at all times, the
renormalized back--reaction tends to $v^2/3$, not $v^2$. It ``stops at
the spinodal line''. The same picture applies for a long time, all
during the ``slow rolling down'' (see section \ref{num}), to
evolutions that start close enough to $\bar\phi=0$. This fact is at the 
basis of the so--called {\em spinodal inflation}\cite{spinflation}.

In any case, the dynamical Maxwell construction, either complete or
partial, poses an interesting question by itself. In fact it is not at
all trivial that the effective potential, in any of the approximation
previously discussed, does bear relevance on the asymptotic behavior
of the infinite--volume system whenever a fixed point is
approached. Strictly speaking in fact, even in such a special case it
is not directly related to the dynamics, since it is obtained from a
static minimization of the total energy at fixed mean field, while the
energy is not at its minimum at the initial time and is exactly
conserved in the evolution. On the other hand, if a solution of the
equations of motion (\ref{Emotion}) exists in which the background
$\bar\phi$ tends to a constant $\bar\phi_\infty$ as $t\to\infty$, one
might expect that the effective action (which however is nonlocal in
time) somehow reduces to a (infinite) multiple of the effective
potential, so that $\bar\phi_\infty$ should be an extremal of the
effective potential. This is still an open question that deserves
further analytic studies and numerical confirmation.

\begin{figure} %7
\epsfig{file=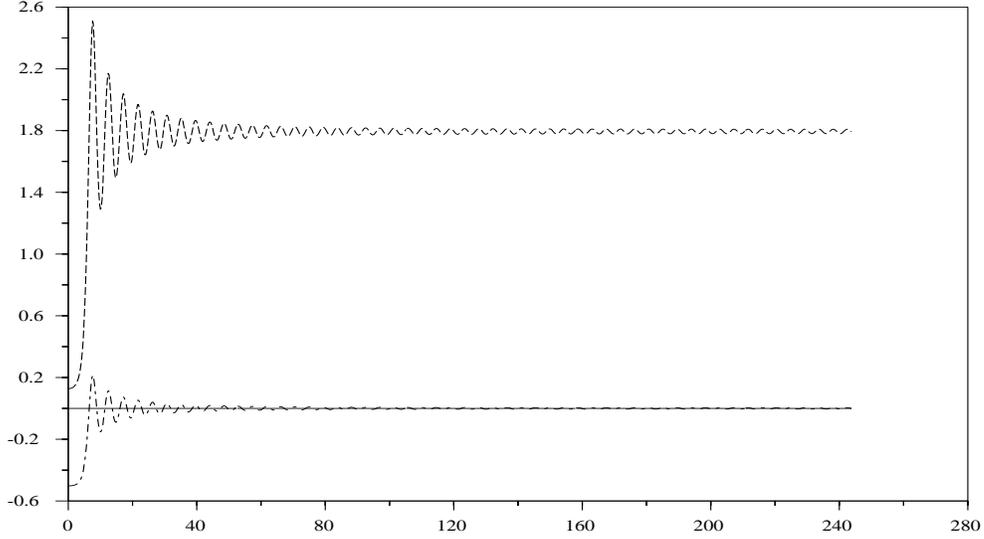,height=8.5cm,width=15cm}
\caption{\it The evolution of the mean value (solid line), the quantum
back--reaction $\Sigma$ (dashed line) and the squared effective mass
$M^2$ (dotted-dashed line), for $\bar\phi=0$ at $t=0$.}
\label{simm:ev}
\end{figure}

\begin{figure} %8
\epsfig{file=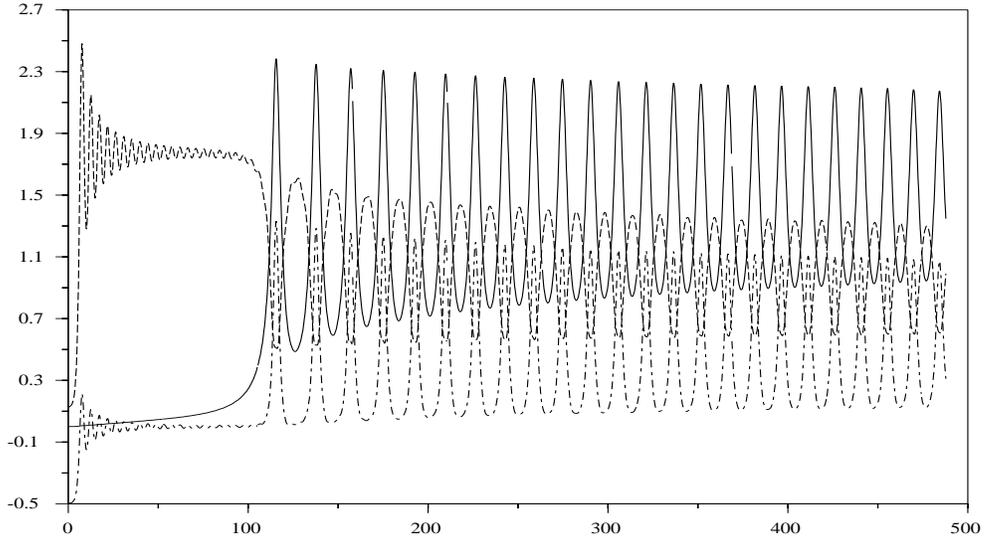,height=8.5cm,width=15cm}
\caption{\it The evolution of the mean value (solid line), the quantum
back--reaction $\Sigma$ (dashed line) and the squared effective mass
$M^2$ (dotted-dashed line), for $\bar\phi=10^{-4}$ at $t=0$, and
$\l=0.1$. The field rolls down very slowly at the beginning.}
\label{slowrd}
\end{figure}

\begin{figure} %9
\epsfig{file=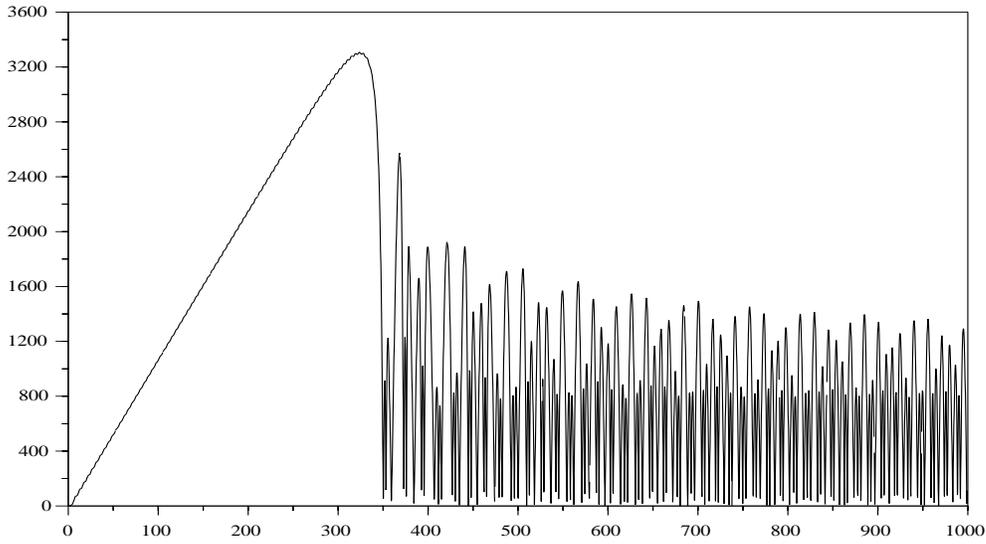,height=8.5cm,width=15cm}
\caption{\it Evolution of the amplitude of the zero mode for $\l=0.1$
and $\bar\phi=10^{-5}$.}
\label{slowrd:0m}
\end{figure}

It is worth noticing also that when the field starts very close to the
top of the potential hill, it remains there for a very long time and
evolves through a very slow rolling down, before beginning a damped
oscillatory motion around a point in the metastability region. During
the slow roll period, $M^2$ oscillates around zero with decreasing
amplitude and the ``phenomenology'' is very similar to the evolution
from symmetric initial conditions, as can be seen comparing Fig.s
\ref{simm:ev} and \ref{slowrd}. Fig. \ref{slowrd:0m} shows
the evolution of the zero mode amplitude in case of a very slow
rolling down. In such a case, after a very short (compared to the time
scale of the figure) period of exponential growth (the spinodal time),
the quantum fluctuations start an almost linear growth, very similar
to the evolution starting from a completely symmetric initial
state. This, obviously, corresponds to the vanishing of the effective
mass. In the meanwhile, $\bar\phi$ keeps growing and rolling down the
potential hill with increasing speed towards the minimum of the
classical potential, eventually entering the metastable region. At
that time, the effective mass starts to increase again and the zero
mode stops its linear growth, turns down and enters a phase of
``wild'' evolution. This time scale, let us call it $\tau_{\text srd}$,
depends on the initial value of the condensate: the smaller
$\bar\phi(t=0)$ is, the longer $\tau_{\text srd}$ will be. We find
numerically that $\tau_{\text srd} \propto \left( \bar\phi(t=0)
\right) ^{1/2}$.

\begin{figure} %10
\epsfig{file=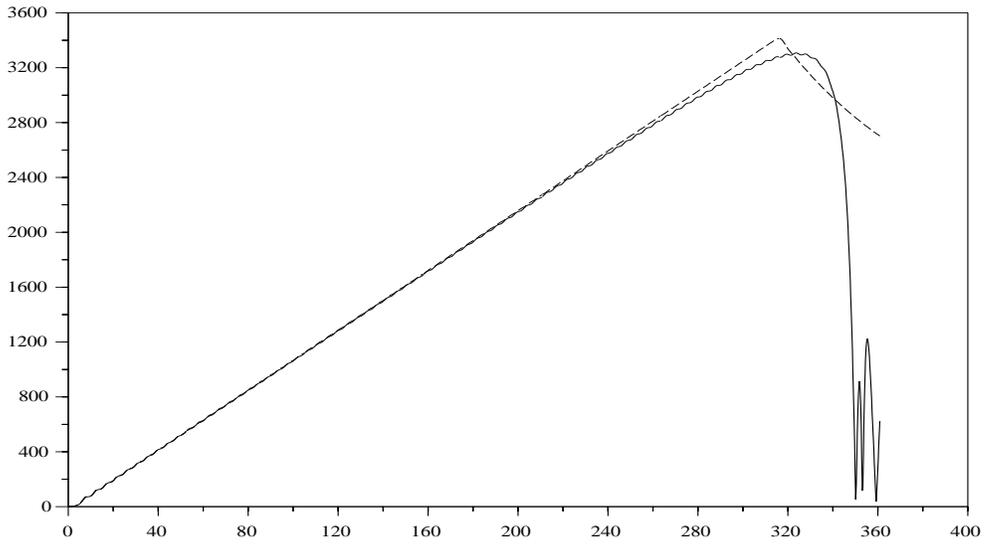,height=8.5cm,width=15cm}
\caption{\it Comparison between the evolutions of the zero mode
amplitude in the following two situations: the dashed line corresponds
to a finite volume simulation with $L/2\pi = 100$ and $\bar\phi=0$,
while the solid line refers to the infinite volume evolution, with
$\bar\phi=10^{-5}$. Both correspond to $\l=0.1$.}
\label{slowrd_vs_fv}
\end{figure}

If we now study the dynamics in finite volume, starting from
condensates different from zero, we will find a competition between
$\tau_{\text srd}$ and $\tau _L$, the time scale characteristic of the
finite volume effects, that is proportional to the linear size of the
box we put the system in. Fig. \ref{slowrd_vs_fv} shows clearly
that when $L/2\pi=100$ and $\bar\phi=10^{-5}$, we have $\tau_{\text
srd} \sim \tau _L$. In any case, either one or the other effect will
prevent the zero mode amplitude from growing to macroscopic values for
any initial condition we may start with. 

It should be noted, also, that the presence of the time scale $\tau _{\text
srd}$ does not solve the internal inconsistency of the gaussian
approximation described above in section \ref{ooed}. In fact, for any
fixed value $L$ for the linear size of the system, we can find a whole
interval of initial conditions for the mean field, which leave enough
time to the fluctuations for growing to order $L$, much before the
field itself had rolled down towards one of the minima of the
classical potential. For those particular evolutions, we would need to
consider the quartic terms in the hamiltonians that the gaussian
approximation neglects, as already explained. 

In addition, there will be also initial conditions for which $\tau _L
> \tau_{\text srd}$. In that case, the effective mass soon starts
oscillating around positive values and it is reasonable to think that
it will take a much longer time than $\tau_L$ for the finite volume
effects to manifest. In \cite{finvN} we have interpreted the
proportionality between $\tau_L$ and $L$ as an auto interference
effect (due to periodic boundary conditions) suffered by a Goldstone
boson wave, traveling at speed of light, at the moment it reaches the
borders of the cubic box. Here, the massless wave we have in the early
phase of the evolution, rapidly acquires a positive mass, as soon as
the condensate rolls down; this decelerates the wave's propagation and
delays the onset of finite volume effects. The gaussian approximation
appears to be fully consistent when we limit ourselves to the
evolution of these particular configurations.

\section{Numerical analysis}\label{num}
We discuss in this section the asymptotic behavior of the dynamical
evolution as it turns out from our numerical results in the gaussian
approximation.

Let us begin with the precise form of the evolution equations for the
field background and the quantum mode widths, as described in sections
\ref{ooed} (cfr eq. (\ref{Emotion}).
\begin{equation}\label{snumeq}
\left[ \frac{d ^2}{dt ^2} + \left(M ^2 - 2 \l \phi^2 \right) \right]
\phi = 0 \;, \quad \left[ \frac{d ^2}{dt ^2} + k_n^2+ M ^2
\right] \s_n - \frac1{4\s_n^3} =0
\end{equation}
where the index $n$ labels the discrete set of values used to
perform the sum (finite volume) or the integral (infinite volume) over
momenta in the quantum back--reaction $\Sigma$, while
$M^2(t)$ is defined by the eq. (\ref{unbtdgap}) in case of unbroken
symmetry and by eq. (\ref{btdgap}) in case of broken symmetry. The
back--reaction $\Sigma$ reads, in the notations of this appendix
\begin{equation*}
\Sigma = \sum_{n=0}^{{\cal N}} g_n \s_n ^2 
\end{equation*}
where $g_n$ is the appropriate ``degeneracy'' factor and $\cal N$ is
the number of modes with distinct dynamics. 
Technically it is simpler to treat an equivalent set of equations,
which are formally linear and do not contain the singular Heisenberg
term $\propto \s_n^{-3}$. This is done by introducing the complex mode
amplitudes $z_n=\s_n\exp(i\t_n)$, where the phases $\t_n$ satisfy
$\s_n^2\dot\t_n=1$. Then we find a discrete version of the equations
studied for instance in ref \cite{devega2}, namely
\begin{equation}\label{numeq}
\left[ \frac{d ^2}{dt ^2} + k_n^2 + M ^2 \right] z_n=0 \;,\quad 
\Sigma = \frac{1}{L^D} \sum_{n=0}^{\cal N} g_n |z_n| ^2
\end{equation}
subject to the Wronskian condition
\begin{equation*}
	z_n\,\dot{\bar{z_n}} - \bar{z_n}\,\dot z_n = -i
\end{equation*}
One realizes that the Heisenberg term in $\s_n$ corresponds to the
centrifugal potential for the motion in the complex plane of
$z_n$. Looking at the figs. \ref{fig:m1} or \ref{slowrd:0m}, we can see that
the motions of the quantum modes correspond qualitatively to orbits
with very large eccentricities. In fact, there are istants in which
$\s_n$ is very little and the angular velocity $\dot\t_n$ is very
large. This is the technical reason for preferring the equations in
the form (\ref{numeq}).

To solve these evolution equations, we have to choose suitable initial
conditions respecting the Wronskian condition. In case of unbroken
symmetry, the requirement of minimum energy for the quantum
fluctuations leads to the massive particle spectrum:
\begin{equation*}
z _n ( 0 ) = \frac{1}{\sqrt{2 \Omega _n}} \hspace{1 cm} \frac{d z _n}{dt} ( 0
) = \imath \sqrt{\frac{\Omega _n}{2}}
\end{equation*}
where $\Omega _n = \sqrt{k ^2 _n + M ^2 ( 0 )}$ and the initial
squared effective mass $M ^2 (t= 0)$, has to be determined
self-consistently, by means of its definition (\ref{unbtdgap}).

In case of broken symmetry, the gap equation is a viable mean for
fixing the initial conditions only when $\phi$ lies outside the
spinodal region [cfr. eq (\ref{bnice})]; otherwise, the gap equation
does not admit a positive solution for the squared effective mass and
we cannot minimize the energy of the fluctuations. 
Following the discussion presented in \ref{ooed}, one possible
choice is to set $\s_k^2 = \frac1{2\sqrt{k^2+|M^2|}}$ for $k^2<|M^2|$
and then solve the corresponding gap equation (\ref{newgap}). We will
call this choice the ``flipped'' initial condition. An other
acceptable choice would be to solve the gap equation, setting a
massless spectrum for all the spinodal modes but the zero mode, which
is started from an arbitrary, albeit microscopic, value. This choice
will be called the ``massless'' initial condition.

\begin{figure} %11
\epsfig{file=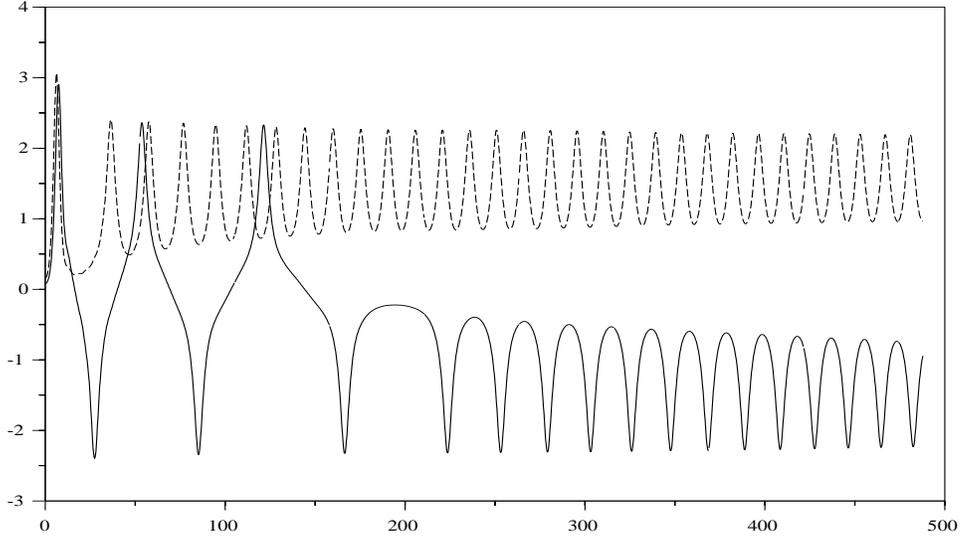,height=8.5cm,width=15cm}
\caption{\it Evolution of the mean value $\bar\phi$ for $\l=0.1$ and
for two different initial conditions: $\bar\phi=0.08$ (solid line) and
$\bar\phi=0.16$ (dashed line), with the ``flipped'' choice for the
spinodal modes.}
\label{boths:1}
\end{figure}

\begin{figure} %13
\epsfig{file=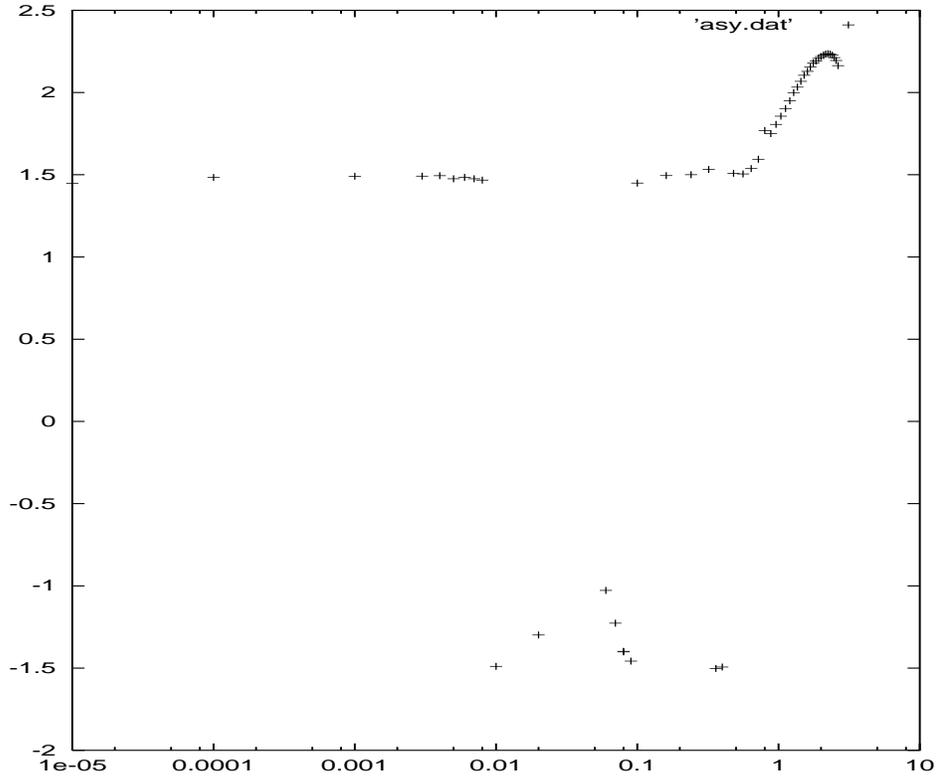,height=10.5cm,width=13cm}
\caption{\it Asymptotic values of the mean field $\bar\phi$, plotted
vs. initial values $\bar\phi(t=0)$, for $\l=0.1$.}
\label{asy:fig}
\end{figure}

\begin{figure} %14
\epsfig{file=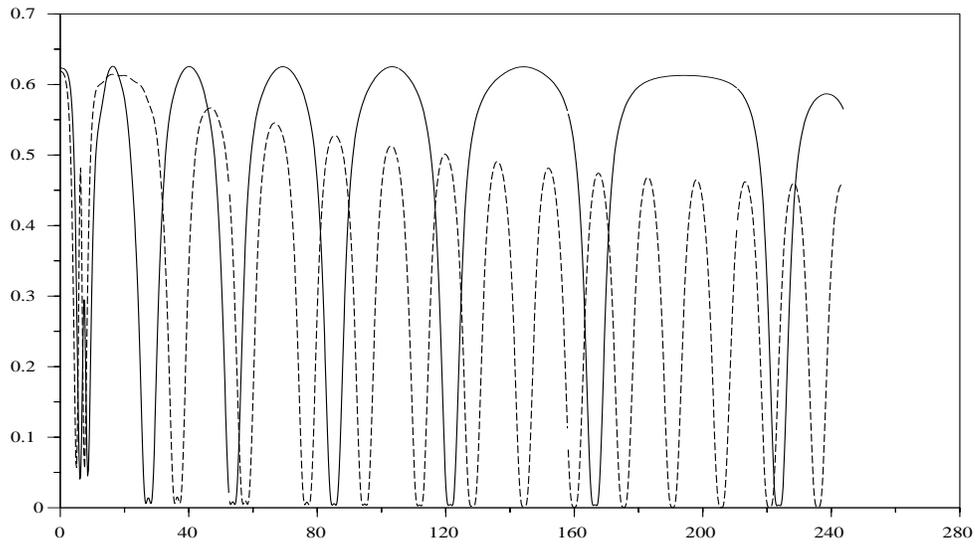,height=8.5cm,width=15cm}
\caption{\it Comparison between the classical energies for the two
initial conditions of Fig \ref{boths:1}.}
\label{energies}
\end{figure}

\begin{figure} %15
\epsfig{file=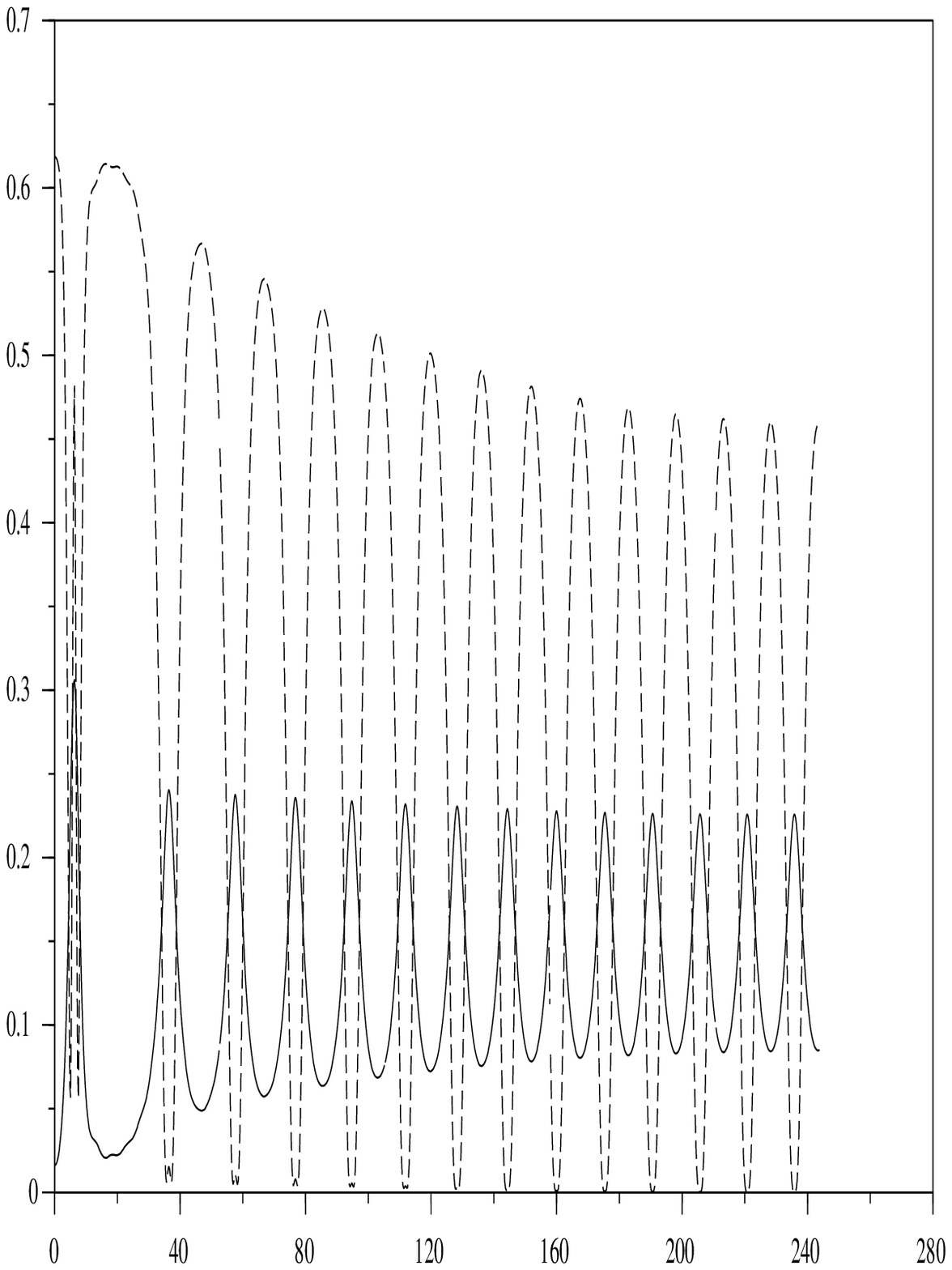,height=8.5cm,width=15cm}
\caption{\it Evolution of the condensate $\bar\phi$ (solid line) and
of the corresponding classical energy (dashed line), for
$\bar\phi(t=0)=0.16$ and $\l=0.1$ (cfr. Fig.s \ref{energies} and
\ref{boths:1}).}
\label{cond_en1}
\end{figure}

\begin{figure} %16
\epsfig{file=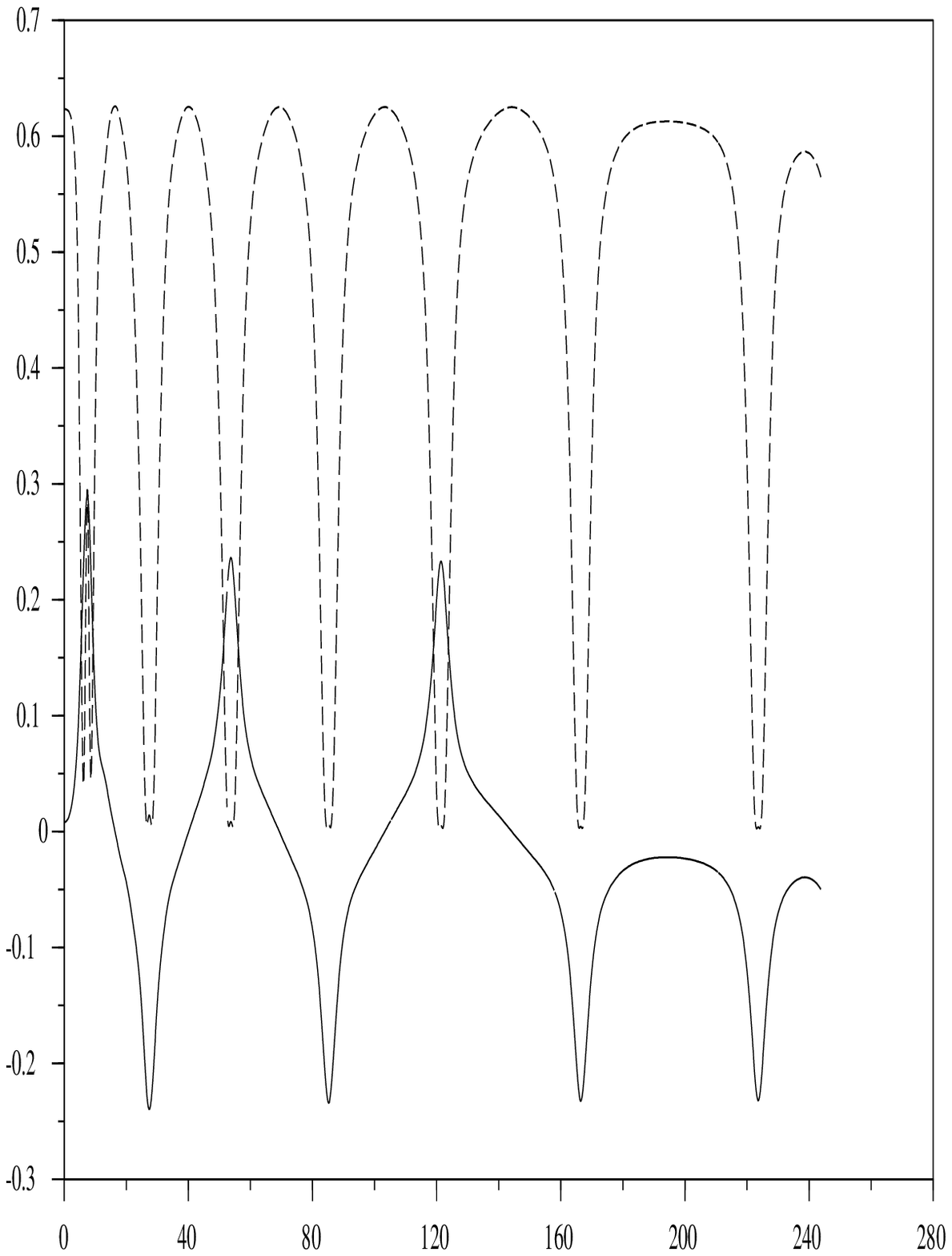,height=8.5cm,width=15cm}
\caption{\it Evolution of the condensate $\bar\phi$ (solid line) and
of the corresponding classical energy (dashed line), for
$\bar\phi(t=0)=0.08$ and $\l=0.1$ (cfr. Fig.s \ref{energies} and
\ref{boths:1}).}
\label{cond_en2}
\end{figure}

Before passing to discuss the influence of different initial
conditions on the results, let us present the asymptotic behavior we
find when we choose the flipped initial condition. In
Fig. \ref{asy:fig} we have plotted the asymptotic values of the mean
field versus the initial values, for $\l=0.1$. All dimensionful
quantities are expressed in terms of the suitable power of the
equilibrium mass $m$. For example, the vev of the field is equal to
$\sqrt{5}$ in these units. First of all, consider the initial values
for the condensate far enough from the top of the potential hill, say
between $\bar\phi(t=0)=0.88$ and $\bar\phi(t=0)=2.64$. In that region
the crosses seem to follow a smooth curve, that has its maximum
exactly at $\bar\phi _{\infty} = \sqrt{5}$ (the point of stable
equilibrium). When we start from an initial condition smaller than
$\bar\phi(t=0)=0.88$, the asymptotic value $\bar\phi _{\infty}$ is not
guaranteed to be positive anymore. On the contrary, it is possible to
choose the initial condition in such a way that the condensate will
oscillate between positive and negative values for a while, before
settling around an asymptotic value near either one or the other
minimum, as fig \ref{boths:1} clearly shows. Fig.s \ref{energies},
\ref{cond_en1} and \ref{cond_en2} helps to understand this behavior by
consideration on the energy balance. Both the evolutions are such that
the classical energy, defined as $\l (\bar\phi ^2 - v^2) /4$, is not a
monotonically decreasing function of time. Indeed, energy is
continuously exchanged between the classical degree of freedom and the
quantum fluctuations bath, in both directions. However, the two rates
of energy exchange are not exactly the same and an effective
dissipation of classical energy on average can be seen, at long time
at least. Of course, this is not the case for the initial transient
part of the evolution starting from the initial condition
$\bar\phi(t=0)=0.08$; there, the condensate absorbs energy (on
average) from the quantum fluctuations, being able to go beyond the
top of the potential hill, towards the negative minimum. This happens
because in case of broken symmetry, the minimization of the
fluctuation energy, within microscopic gaussian states, is not
possible for initial conditions in the spinodal region [cfr. the
discussion about the gap equation (\ref{bnice}) in section
\ref{ooed}]. After a number of oscillations, the energy starts to flow
from the condensate to the quantum bath again (on the average),
constraining the condensate to oscillate around a value close to one
of the two minima. If we look at fig. \ref{asy:fig} again, we can find
positive asymptotic values as well as negative ones, without a
definite pattern, in the whole interval $[0.01,0.8]$. If we start with
$0<\bar\phi(t=0)<0.01$ we have the slow rolling down, already
described in section \ref{ep} and the mean field oscillates around a
positive value from the beginning, never reaching negative values. A
further note is worth being added here. During the phase of slow
rolling down, the evolution is very similar to a symmetric evolution
starting from $\bar\phi(t=0)=0$; in that case, the dissipation
mechanism works through the emission of (quasi-)massless particles and
it is very efficient because it has not any perturbative threshold. If
the field stays in this slow rolling down phase for a time long
enough, it will not be able to absorb the sufficient energy to pass to
the other side ever again and it will be confined in the positive
valley for ever. Evidently, when $\bar\phi(t=0)>0.01$ this dissipative
process might not be so efficient to prevent the mean field from
sampling also the other valley.

\begin{figure} %12
\epsfig{file=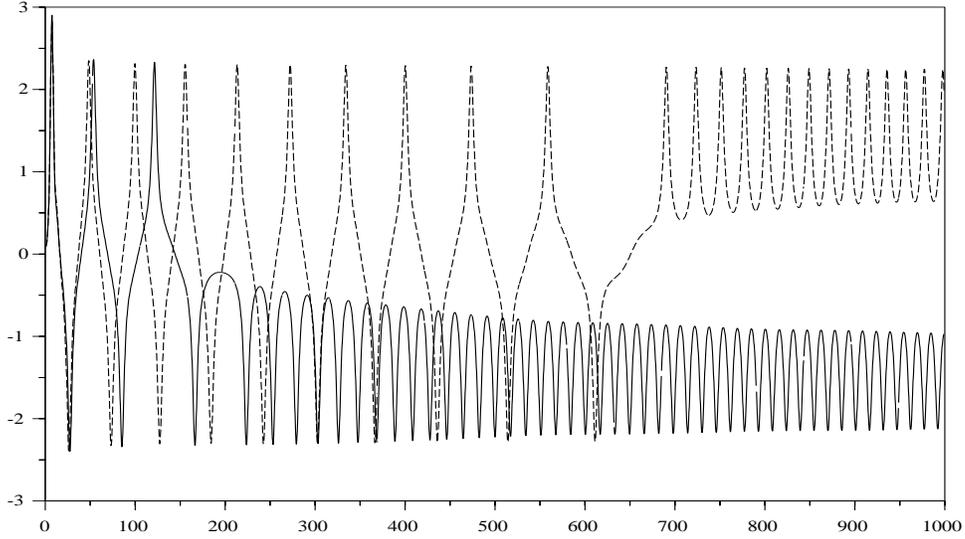,height=8.5cm,width=15cm}
\caption{\it Evolution of the mean value $\bar\phi$ for $\l=0.1$, with
$\bar\phi(t=0)=0.08$, and two different initial conditions for the
quantum spinodal modes, ``flipped'' (solid line) and massless (dashed line).}
\label{boths:2}
\end{figure}

Which one of the two valleys will be chosen by the condensate is a
matter of initial conditions and it is very dependent from the energy
stored in the initial state, as is shown in fig. \ref{boths:2}, where
two evolutions are compared, starting from the same value for the
condensate, but with the two initial conditions, ``flipped'' and
massless, for the quantum fluctuations.

\section{Conclusions and Perspectives}\label{conclusion}
In this work we have extended the standard time dependent Hartree-Fock
approximation \cite{tdHF} for the $\phi^4$ QFT, to include some
non-gaussian features of the complete theory. We have presented a
rather detailed study of the dynamical evolution out of equilibrium,
in finite volume (a cubic box of size $L$ in $3$D), as well as in
infinite volume. For comparison, we have also analyzed some static
characteristics of the theory both in unbroken and broken symmetry
phases.

By means of a proper substitution of the bare coupling constant with
the renormalized coupling constant (fully justified by diagrammatic
consideration), we have been able to obtain equations of motion
completely independent of the ultraviolet cut-off (apart from a slight
dependence on inverse powers, that is, however, ineluctable because of
the Landau pole). We have described in detail the shape of the ground
state, showing how a broken symmetry scenario can be recovered from
the quantum mechanical model, when the volume diverges. 

Moreover, we have shown that, within this slightly enlarged tdHF
approach that allows for non--gaussian wavefunctions, one might
recover the usual gaussian HF approximation in a more controlled way.
In fact, studying the late time dynamics, we have confirmed the
presence of a time scale $\tau_L$, proportional to the linear size $L$
of the box, at which the evolution ceases to be similar to the
infinite volume one. At the same time, the low--lying modes amplitudes
have grown to order $L$. The same phenomenon has been observed in the
$O(N)$ model \cite{finvN}. Looking at this result in the framework of
our extended tdHF approximation, one realizes that the growth of
long--wavelength fluctuations to order $L$ in fact undermines the
self--consistency of the gaussian HF itself. In fact, in our tdHF
approach the initial gaussian wavefunctions are allowed to evolve into
non--gaussian forms, but they simply do not do it in a macroscopic
way, within a further harmonic approximation for the evolution, so
that in the infinite--volume limit they are indistinguishable from
gaussians at all times. But when $M^2$ is on average not or order
$L^0$, but much less, as it happens for suitable initial conditions,
infrared modes of order $L$ will be dominated by the quartic term in
our Schroedinger equations (\ref{Schroedinger}), showing a possible
internal inconsistency of the gaussians approximation.

An other manifestation of the weakness of the HF scheme is the curious
``stopping at the spinodal line'' of the width of the gaussian quantum
fluctuations, when the initial configuration does not break the
symmetry. This does not happen in the large $N$ approach because of
different coupling of transverse mode (the only ones that survive in
the $N\to\infty$ limit) with respect to the longitudinal modes of the
$N=1$ case in the HF approach.

We have also described the non--trivial phenomenology of the
infinite--volume late--time evolution in the gaussian approximation,
showing how the dynamical Maxwell construction differs from the
$N=\infty$ case. In fact, we have observed the presence of an unstable
interval, contained in the static flat region which is forbidden as
attractor of the asymptotic evolution. This region corresponds, more
or less, to the spinodal region of the classical potential, with the
obvious exception of the origin. In particular, we have found that the
energy flux between the classical degree of freedom and the bath of
quantum fluctuations is quite complex and not monotonous. In other
words, since we start from initial conditions where the fluctuation
energy is not minimal, there are special situations where enough
energy is transferred from the bath to the condensate, pushing it
beyond the top of the potential hill. 

Clearly further study, both analytical and numerical, is needed in our
tdHF approach to better understand the dynamical evolution of quantum
fluctuations in the broken symmetry phase coupled to the condensate.
An interesting direction is the investigation of the case of finite
$N$, in order to interpolate smoothly the results for $N=1$ to those
of the $1/N$ approach.  It should be noted, in fact, that the theory
with a single scalar field contains only the longitudinal mode (by
definition), while only the transverse modes are relevant in the large
$N$ limit.  Hence a better understanding of the coupling
between longitudinal and transverse modes is necessary.

In this direction, another relevant point is whether the Goldstone
theorem is respected in the HF approximation
\cite{onhartree}. It would be interesting also to study the dynamical
realization of the Goldstone paradigm, namely the asymptotic vanishing
of the effective mass in the broken symmetry phases, in different
models; this issue needs further study in the $2D$ case \cite{chkm},
where it is known that the Goldstone theorem is not valid.

\section{Acknowledgements}
C. D. thanks D. Boyanovsky, H. de Vega, R. Holman and M. Simionato
for very interesting discussions. C. D. and E. M. thank MURST and INFN for
financial support.

\vskip 0.5 truecm

\vskip 1 truecm

%\begin{figure}
%\epsfig{file=mxw.eps,height=8.5cm,width=15cm}
%\caption{\it The static and dynamical Maxwell construction for $m^2=1$
%and $\l=0.1$.}\label{mxw}
%\end{figure}

%
%\begin{figure}
%\epsfig{file=negasy.eps,height=8.5cm,width=15cm}
%\caption{\it Negative asymptotic values of the mean field $\bar\phi$,
%plotted vs. initial values $\bar\phi(t=0)$.}
%\label{}
%\end{figure}
%

\end{document}